\documentclass[a4paper,11pt]{article}
\pdfoutput=1 
\usepackage{jheppub} 

\usepackage{amsmath,amscd}
\usepackage{amsfonts}
\usepackage{amssymb}
\usepackage{graphicx,shortvrb}
\usepackage{dsfont}

\usepackage[utf8]{inputenc}

\def\be{\begin{equation}}
\def\ee{\end{equation}}
\def\bea{\begin{eqnarray}}
\def\eea{\end{eqnarray}}

\def\pa{\partial}
\def\to{\rightarrow}
\def\nonu{\nonumber \\{}}

\def\cald{{\cal D}}

\def\calh{{\cal H}}

\def\call{{\cal L}}

\def\caln{{\cal N}}
\def\calo{{\cal O}}

\def\cals{{\cal S}}

\def\a{\alpha}
\def\b{\beta}
\def\g{\gamma}
\def\G{\Gamma}
\def\d{\delta}
\def\e{\epsilon}

\def\D{\Delta}
\def\h{\eta}
\def\l{\lambda}
\def\L{\Lambda}
\def\k{\kappa}

\def\F{\Phi}

\def\m{\mu}
\def\n{\nu}

\def\O{\Omega}
\def\p{\pi}
\def\r{\rho}

\def\s{\sigma}
\def\S{\Sigma}

\def\x{\xi}
\def\half{{1\over 2}}

\title{Superconformal mechanics of AdS$_2$ D-brane boundstates} 
\author[a]{Delaram Mirfendereski,}
\author[b]{Joris Raeymaekers}
\author[a,c]{and Dieter Van den Bleeken}
\affiliation[a]{Physics Department, Boğaziçi University\\
		34342 Bebek / Istanbul, TURKEY}
\affiliation[b]{CEICO, Institute of Physics of the ASCR, \\ Na Slovance 2, 182 21 Prague 8, Czech Republic}
\affiliation[c]{Secondary address:\\
		Institute for Theoretical Physics, KU Leuven\\
		3001 Leuven, Belgium}
		
\emailAdd{delaram.mirfendereski@boun.edu.tr}
\emailAdd{joris@fzu.cz}
\emailAdd{dieter.van@boun.edu.tr }
		
\abstract{
	We explicitly construct a family of ${\cal N}=4$ superconformal mechanics of dyonic particles, generalizing the work of Anninos et al.\,\cite{Anninos:2013nra}  to an arbitrary number of particles. These mechanics are obtained from a scaling limit of the effective Coulomb branch description of ${\mathcal N}=4$ quiver quantum mechanics describing D-branes in type II Calabi-Yau compactifications. In the supergravity description of these D-branes this limit changes the asymptotics to AdS$_2\times $S$^2\times$CY$_3$. We exhibit the $D(1,2;0)$ superconformal symmetry and conserved charges of the mechanics in detail. In addition we present an alternative formulation as a sigma model on a hyperk\"ahler manifold with torsion.}
	
\arxivnumber{}

\begin{document}
\maketitle
\section{Introduction}
Similarly to BPS field theory solitons \cite{Manton:1981mp}, extremal black holes can be studied in the moduli space approximation \cite{Ferrell:1987gf}. When embedded in supergravity this gives rise to a supersymmetric multi-particle mechanics that in an appropriate limit becomes also conformally invariant  \cite{Michelson:1999dx, Maloney:1999dv}, see \cite{BrittoPacumio:1999ax} for a review. Such models are of interest due to their potential relation to the microscopics of black hole entropy and AdS$_2$/CFT$_1$ duality. The original work at the end of  the 90's focused on equally charged black holes. In the early 2000's it was realized that in addition there also exists a large class of interacting  BPS black holes with mutually non-local charges, leading to non-trivial bound states \cite{Denef:2000nb, LopesCardoso:2000qm}. From a 4d perspective they are the gravitational backreaction of dyonically charged point particles that originate from wrapped D-branes in type II string theory Calabi-Yau compactifications. The supersymmetric (quantum) mechanics of these dyonic particles, including some stringy interactions, was derived in \cite{Denef:2002ru} and takes the form of a 1d, {$\cal N$}=4 supersymmetric quiver gauge theory. Integrating out the stringy modes when the gauge theory  is in its Coulomb phase reduces it to the mechanics of $N$ point particles, whose essential features can be identified with those of $N$ dyonic BPS black holes in $\caln=2$ 4d supergravity, due to a powerful non-renormalization theorem \cite{Denef:2002ru}.  The physics of these BPS bound states and their stability has developed into a wide area of study but in this paper we will focus only on some aspects of a special class of bound states known as {\it scaling solutions} \cite{Denef:2007vg, Bena:2012hf}, where the dyonic centers can approach each other arbitrarily closely. It was shown in \cite{Anninos:2013nra} that for three such scaling centers there exists a limit where the Coulomb quiver mechanics becomes superconformally invariant and, as we detail in this paper, this can be extended to an arbitrary number of scaling centers. We find that the superconformal algebra governing this limit is $D(2,1;0)$, which also appears in the moduli space dynamics of marginally bound black holes \cite{Maloney:1999dv}.

On the supergravity side the limit produces a number of dyonic BPS black holes in an asymptotic AdS$_2\times$S$^2$ space-time \cite{Bena:2018bbd,Mirfendereski:2018tob}. The $N$-particle superconformal mechanics we present in this paper thus captures some of the physics of asymptotically AdS$_2$ black hole bound states. 
We hope that our results pave the way, upon quantizaton, towards a more concrete connection between D-brane state counting,  AdS$_2$/CFT$_1$ \cite{Sen:2008yk} and black hole microscopics. In particular the conformal scaling regime appears to sit at the crucial supergravity-Coulomb-Higgs trisection and could hence help clarify how much of the black hole microscopics one can expect to be captured by supergravity. 

Parallel to the advances in understanding the BPS sector of $\caln=2$ 4d supergravity/string theory since the first work on superconformal black hole mechanics, there has also been a lot of progress in the framework underlying ${\cal N}=4$ supersymmetric (quantum) mechanics, see e.g. \cite{Ivanov:2011gk, Fedoruk:2011aa} for reviews. Some of the most important advances concern the superspace formulation, dualities between different multiplets and a more thorough understanding of the underlying geometry. Although by now there exists a wide and detailed overview of the generic features of these models, the number of non-trivial and physically relevant examples remains somewhat scarce. The $N$ particle models studied in this paper add a rich new class of explicit examples. The quiver mechanics we discuss was originally, and most naturally, phrased in terms of the so-called $\mathbf{(3,4,1)}$ multiplet\footnote{The bracket notation denotes the number of {\bf (bosonic, fermionic, auxiliary)} fields in the multiplet.}  and this is also the language we use in most of our paper. But in addition we work out explicitly the reformulation in terms of $\mathbf{(4,4,0)}$ multiplets, which provides a more powerful geometric interpretation in terms of a hyperk\"ahler with torsion (HKT) sigma model. This is an important step towards quantization which in the HKT formulation can be done in terms of differential forms \cite{Smilga:2012wy}.

\subsection{Outline and overview of results}
We start the paper in section \ref{CQsec} presenting the supersymmetric mechanics of $N$ dyonic particles, originating in the effective Coulomb branch description of Denef's ${\cal N}=4$ quiver quantum mechanics \cite{Denef:2002ru}. We introduce the model in $\caln=4$ superspace and re-derive its component form. One of the main aims of this paper is to provide a detailed understanding of the full symmetry and conserved charges of the family of models under consideration, and for this reason we carefully keep track of these from the start. Apart from providing details not previously spelled out in the literature, this section also introduces one novelty. All non-trivial physics finds itself in the relative dynamics of the dyonic particles and it is only after decoupling the center of mass that a conformal limit can be taken. We thus present only the Lagrangian for the relative dynamics, but instead of doing so by choosing $3N-3$ (complicated) adapted coordinates we keep working with the full set of $3N$ positions of all particles such that the overall translations appear as a gauge symmetry. This trick allows us to extend \cite{Anninos:2013nra} from three to an arbitrary number of centers, but it also has some subtle consequences. For example, the metric appearing in the kinetic term becomes semi-positive definite, but we show how using a projective inverse the standard geometric intuition behind the component Lagrangian can be kept intact.

In section \ref{confsec} we show how upon taking an appropriate limit the supersymmetric mechanics of section \ref{CQsec} gains superconformal symmetry, extending \cite{Anninos:2013nra} to an arbitrary number of centers. We exhibit the full $D(2,1;0)$ superconformal symmetry and corresponding conserved charges. As a first step towards quantization we reconsider our models in the Hamiltonian formalism in section \ref{Hamsec}. We introduce canonical variables and derive the $D(2,1;0)$ algebra in terms of Poisson brackets.

In section \ref{HKTsec} we reformulate the supersymmetric mechanics in terms of the $\mathbf{(4,4,0)}$ multiplet, instead of the $\mathbf{(3,4,1)}$ multiplet used in the previous sections. We write out the precise HKT geometry underlying the sigma model that appears in this form and comment on how the reformulation can be understood in terms of gauging an extra direction, thus providing an example of the general notion of automorphic duality as discussed in \cite{Delduc:2006yp}. Also in HKT form our mechanics are conformally invariant but, as we point out, the conformal transformations differ slightly from the 'standard' ones of \cite{Michelson:1999zf}.

We end the paper in section \ref{discsec} by discussing some of the physical properties of the mechanical models we developed in the previous sections and comment on possible future directions. 

There are also 5 appendices collecting some relevant technicalities.


\section{Review of Coulomb branch  quiver mechanics}\label{CQsec}
We will study a class of  `quiver' mechanical systems  describing the dynamics of $N$ D-branes in type II string theory which are wrapped on  internal Calabi-Yau cycles and which are pointlike in the three noncompact spatial directions. When the D-brane centers are sufficiently far apart, the fundamental strings stretching between them can be integrated out to give an effective  (`Coulomb branch') mechanics for the $N$ 3d position vectors of the centers and their superpartners.  In this section  we review the Coulomb branch mechanics, which was derived in \cite{Denef:2002ru, Anninos:2013nra}, and its invariance under $\caln = 4$ supersymmetry. Other relevant literature on this type of models includes \cite{Smilga:1986rb, Ivanov:1990jn, Diaconescu:1997ut, Ivanov:2002pc}. 

\subsection{Coulomb branch Lagrangian and symmetries}


The Coulomb branch quiver quantum mechanics for an $N$ centered D-brane system preserving four  supersymmetries describes $N$ interacting vector multiplets.
The off-shell $d=1,\ \caln =4$ vector multiplet can be obtained from the $d=4,\ \caln =1$ vector multiplet by dimensional reduction  \cite{Smilga:1986rb}. The bosonic content consists of three spatial coordinates ${x_i},i = 1, 2 ,3$, a worldline gauge potential $C_t$ and an auxiliary field $D$. The fermionic superpartners form a  2-component spinor $\l_\a, \a= 1,2$ and its complex conjugate $\bar \l^\a \equiv (\l_\a)^*$. 
Assuming that all nodes in the quiver are $U(1)$ -- which is equivalent to assuming all brane/center charges to be primitive -- the gauge field does not couple to the other vector multiplet fields, since they are valued in the adjoint representation which is trivial for $U(1)$\footnote{In case the nodes are not primitive the vector multiplet can be $U(M)$ valued and this will lead to a more complicated Lagrangian, see e.g. $L_V$ on page 41 of \cite{Denef:2002ru}, where $\cald_t$ involves the connection $C_t$ ($A$ in \cite{Denef:2002ru}'s nomenclature.)}. So after the hypermultiplets are integrated out the wordline gauge potentials $C_t$ are completely absent from the effective vector multiplet Lagrangian which contains only the fields $x^i, \l_\a, \bar \l^\a, D$. These are said to constitute the ${\bf (3, 4, 1)}$ multiplet\footnote{See \cite{Pashnev:2000ij} for the classification of supermultiplets in quantum mechanics.} and  the Lagrangian contains $N$ of them, one for each center.  Equivalent formulations in terms of different multiplets are possible (see e.g. \cite{Ivanov:2003tm}), and in particular the formulation in terms of the ${\bf (4, 4, 0)}$ multiplet will be explored in section \ref{HKTsec}.

\subsubsection{Superspace formulation}
The Coulomb branch quiver quantum mechanics can be formulated in the standard\footnote{There also exists a powerful formulation in terms of harmonic superspace, see e.g. \cite{Ivanov:2003nk, Ivanov:2011gk}, but such machinery is not needed for our purposes.} $\caln = 4$ superspace. While somewhat abstract, the discussion of the symmetries of the model is most economical in this formulation and therefore we review it here and in Appendix \ref{Appsusy} -- see \cite{Diaconescu:1997ut, Ivanov:2002pc} for further details and references --, while for the discussion of the Noether charges and algebra  we will work in the more transparant component formalism.

The $d=1,\ \caln =4$ superspace $\mathbb{R}^{1|4}$ is parametrized by the time $t$ and an anticommuting $SU(2)$ spinor doublet $\theta_\a$ together with its  complex conjugate $\bar \theta^\a = (\theta_\a)^*$. Our spinor conventions are spelled out in Appendix \ref{Appspinors}.
The $\caln = 4$ superspace admits an action of the $\caln = 4$ Poincar\'e supersymmetry
algebra  generated by
\be
H = \pa_t, \qquad Q^\a = {\pa \over \pa \theta_\a} + i \bar \theta^\a \pa_t, \qquad \overline Q_\a =  {\pa \over \pa \bar \theta^\a} + i  \theta_\a \pa_t.
\ee
These satisfy the anticommutation relations
\be \{ Q^\a, \bar Q_\b \} = 2 i \d^\a_\b H. \label{Poincsusy}
\ee
Also relevant for this work will be the action of two commuting $su(2)$ algebras which will play the role of R-symmetries. Under the first  $su(2)$, the $\theta_\a$ and the $\bar \theta^\a$ transform as doublets, while the second 
algebra (denoted henceforth as $\widetilde{su(2)}$) mixes the $\theta$ and $\bar \theta$ coordinates: \begin{align}
R_i =& {i \over 2} (\s_i \theta)_\a { \pa \over \pa {\theta_\a}}-{i \over 2} (\bar \theta \s_i )^\a { \pa \over \pa {\bar \theta^\a}},&&&&\\
\tilde R_3 =&-{i \over 2} \left( \theta_\a {\pa \over \pa \theta_\a}-\bar\theta^\a {\pa \over \pa \bar \theta^\a}\right), &\tilde R_+ =& (\e \theta)^\a {\pa \over \pa \bar \theta\a}, & 
\tilde R_- = & (\bar \theta \e)_\a {\pa \over \pa  \theta_\a}, 
\end{align}
where $\tilde R_\pm \equiv \tilde R_1 \pm i \tilde R_2 $. 
These lead to the additional  nonvanishing commutators
\begin{align}
[ R_i , R_j ] =& \e_{ijk}R_k, & [\tilde R_i , \tilde R_j ] =& \e_{ijk}\tilde R_k, &
[ R_i , Q^\a ] =&- {i \over 2} ( Q\s_i)^\a, & [ R_i ,\overline Q_\a ] =& {i \over 2} (\s_i \overline Q )_\a\nonu
[\tilde R_+ , Q^\a ] =& (\e \overline Q)^\a, & [\tilde R_- , \overline Q^\a ] =&- (\e  Q)_\a, &
[\tilde R_3 , Q^\a ] =&{i \over 2} Q^\a, & [\tilde R_3 , \overline Q_\a ] =&- {i \over 2} \overline Q_\a.\label{Rsymm}
\end{align}
The supercovariant derivatives which anticommute with $Q^\a$ and $\overline Q_\a$ are defined as
\bea
D^\a &=& {\pa \over \pa \theta_\a}- i \bar \theta^\a \pa_t, \qquad \bar D_\a =  {\pa \over \pa \bar \theta^\a}  - i  \theta_\a \pa_t
\eea
and satisfy 
\be
\{ D^\a, \overline D_\b \} =- 2 i \d^\a_\b \pa_t.
\ee

The off-shell vector multiplet can be described by a real scalar superfield 
\be 
V = V^*,
\ee
subject to the gauge equivalence
\be 
V \sim V + \L + \bar \L\label{Vgauge}
\ee 
where $\L$ is a chiral superfield with $ \overline D_\a \L= D^\a \bar \L=0$.
The field $V$ transforms as a scalar both under  Poincar\'e supersymmetry and $R$-symmetry, i.e.
\be 
\d V = (u H +  \xi Q + \bar \xi \bar Q + r^i R_i + \tilde r^i \tilde R_i) V.\label{Vsusy}
\ee
Here, $u, \x, \bar \x, r^i$ and $\tilde r^i$ denote the infinitesimal parameters for time translation, supersymmetry and the $R$-symmetries  respectively. 

 From $V$ we can form a triplet of gauge-invariant superfields
\be
\F_i = \half D \s_i   \bar D V 
, \qquad i = 1,2,3.\label{FitoV}
\ee
From (\ref{Vsusy}) and (\ref{FitoV}) one finds, by commuting the superspace  generators through the operator $D \s_i   \bar D$,  that the $\F_i$ are scalars under supersymmetry, transform as a triplet under $su(2)$ and as a singlet under $\widetilde{su(2)}$, i.e.
\be
\d  \F_i = \e_{ijk} r^j \F^k + (u H+ \xi Q + \bar \xi \bar Q +  r^j R_j + \tilde r^i \tilde R_i)  \F_i .\label{susyPhi}
\ee
A convenient component parametrization for $V$ is given in (\ref{appVcomp}). The component fields $x^i, \l_\a, \bar \l^\a, D$ introduced at the beginning of this section enter in the superspace expansion of $\F_i$ as  
\be
\F_i = x_i + i \bar \l \s_i \theta - i \bar \theta \s_i \l+ \bar \theta \s_i \theta D- \e_{ijk}\bar \theta \s_j \theta \dot x_k +\half \bar \theta \e \bar \theta \theta \e \s_i \dot\l -\half  \theta \e \theta \bar \theta  \s_i \e \dot{\bar \l} +{1\over 4} \theta \e \theta  \bar \theta \e \bar \theta \ddot x_i.\label{Phicomp}
\ee
The superfields $\F_i$ satisfy a number of constraints as a consequence of (\ref{FitoV}) which are listed in (\ref{appconstr1}, \ref{appconstr2}, \ref{WZWprop1}, \ref{WZWprop2}).

After these preliminaries we are ready to introduce the Lagrangian governing the general Coulomb branch quiver mechanics. To describe an $N$-centered system we
consider superfields $V^a, \F^a_i$ labelled by an  additional index $a= 1, \ldots , N$.
The Coulomb branch quiver Lagrangian splits into two decoupled parts. There is a trivial   universal  term decribing the free motion of the center of mass coordinate of the system and it's superpartners, and a second part capturing the dynamics of the relative motion of the branes.
We will restrict attention to the relative Lagrangian in what follows. In the formulation with off-shell susy, this Lagrangian  admits an expansion in powers of velocities; that is, $L^{(n)}$, the $N$-th term in the expansion is the $\caln=4$ supersymmetric completion of a bosonic term of order $N$ in time derivatives. We will here work up to quadratic order in velocities.  
The terms in the relative Lagrangian can be written as  superspace integrals,
\footnote{We normalize  the superspace measure such that
	\be
	L = \int d^4\theta  F (\F) \equiv - \left. F(\F)\right|_{\theta \e \theta  \bar \theta \e \bar \theta}= {1 \over 4} \left. F(\F)\right|_{\theta_1 \theta_2 \bar \theta^1 \bar \theta^2} .\nonumber
	\ee} 
\bea
L &=&  \int d^4 \theta \left(  \call^{(0)}+ \call^{(1)}+ \call^{(2)} \right) + L_{\rm   bdy}\label{Lsuspace1}\nonu
\call^{(0)} &=& - 2    f_a  V^a \nonu
\call^{(1)} &=& - 2   \int_0^1 dy\, U_a \left(\F^b (y) \right)\pa_y V^a (y)\nonu
\call^{(2)} &=& \calh (\F^a ).\label{Lsuspace4}
\eea
 We have included  a total derivative term $L_{\rm bdy}$ to remove second order time derivatives proportional to $\ddot x^{ia}$  and to guarantee a good variational principle. The explicit form is given below in (\ref{Lbdy}).

The lowest order term  $ L^{(0)}$ is a standard Fayet-Iliopoulos (FI) term. It is gauge-invariant under (\ref{Vgauge}) despite its dependence on the $V^a$.  The constants $f_a$ are the FI parameters which  satisfy, in the models of interest,
\be
\sum_a f_a =0.\label{sumf}
\ee

The first order term $ L^{(1)}$ requires more explanation. Its superspace form was first found in \cite{Ivanov:2002pc} and  resembles a  Wess-Zumino-Witten term. The fields in the integrand depend on an extra parameter $y$ such that $V^a(1) = V^a$ and $V^a (0)$ is a constant. $ L^{(1)}$ is gauge invariant under  (\ref{Vgauge}) and
has  the property that its variation only depends on the fields at $y=1$,
since the variation of the integrand is a total $y$-derivative, namely
\be 
\d L^{(1)} =  - 2  \int d^4 \theta  U_a  \d V^a . \label{L1var}
\ee
These properties severely constrain the functions $U_a$. In our model they are of the form
\be 
U_a = \sum_{b , b\neq a}
{\k_{ab} \over 2 |\F^{ab}|}, \qquad \F^{iab} \equiv \F^{ia}-\F^{ib},\label{Uquiver}
\ee
where 
\be 
\k_{ab} = - \k_{ba}.\label{kappaas}
\ee
Gauge invariance of $ L^{(1)}$  follows from writing the gauge parameter as $\L =\bar D \e \bar D \O$ and using the identity (\ref{WZWprop1}), while the property  (\ref{L1var}) follows from (\ref{WZWprop2}).
The constraints on the form of $U_a$ lie at the basis of an important non-renormalization theorem for $L^{(1)}$ \cite{Denef:2002ru}. 

Lastly,  the second order term in $L$ is the superspace integral of a potential function $\calh$ of the gauge-invariant superfields. 
 In the quiver mechanics of interest this potential takes the form
\be 
\calh = \sum_{a, b , a\neq b}
\left({\m_{ab} |\F^{ab}|^2\over 6}  -\frac{|\kappa_{ab}|}{4|\F^{ab}|}\log |\F^{ab}|\right)\label{Hnonscaling}
\ee
where 
\be 
\m_{ab} =  \m_{ba}\label{muas}.
\ee
We should remark that the potential $\calh$ is not unique and we can add terms to it whose superspace integral yields a total derivative. For example, from the identity (\ref{WZWprop1}) one derives that
\be
\int d^4 \theta {1 \over |\F|} = - {1 \over 4} {d^2 \over dt^2}\left({1 \over |x|}\right),\label{int1F}
 \ee
and therefore we are free to
 add to $\calh$ terms of the form
\be 
\calh \to \calh + \sum_{a, b , a\neq b}
{\l_{ab} \over |\F^{ab}|}\label{Hesstransf}
\ee
for arbitrary constants $\l_{ab}$.

We have now specified the models of interest up to the constants $f_a, \k_{ab}, \m_{ab}$ which are determined by the underlying D-brane physics: the FI parameters $f_a$ are related to the Calabi-Yau moduli of the string compactification, the $\k_{ab}$ are the Dirac-Schwinger-Zwanziger (DSZ) inner products of the D-brane charges of the   centers labelled by $a$ and $b$, and the  $\m_{ab}$ determined by the masses of the centers. We refer to \cite{Denef:2002ru} for more details on the meaning of these parameters.

It is straightforward to see that the action is invariant under all the symmetries advertised above. Since the superspace Lagrangian density $\call$  depends only on  rotationally invariant combinations of the $\F^{ia}$, it transforms as a scalar under all the symmetries,
\be 
\d \call = \left(u H +  \xi Q + \bar \xi \bar Q + r^i R_i + \tilde r^i \tilde R_i\right)\call.
\ee
From the form of the generators given above we see that all the terms in this expression are total derivatives with respect to either time or the anticommuting superspace coordinates.


Though it is not manifest in the way we have written it, the relative Lagrangian $L$ depends only on the superfield differences $V^{ab} = V^a- V^b$ 
and hence only on $N-1$ independent superfields. In other words, it possesses a gauge invariance  under an overall shift by an arbitrary superfield,
\be 
\d_{\rm shift} V^a (t) = v^a \S (t).\label{shiftsf}
\ee
where we defined
\be 
(v^a) = (1,1,\ldots, 1).
\ee
Invariance of $L$ under \eqref{shiftsf} follows
from the properties
\be
0=f_a v^a = U_a v^a= {\pa \calh  \over \pa \F^{i a}}   v^a.\label{shiftss}
\ee
which in turn follow from (\ref{sumf}, \ref{kappaas}, \ref{muas}).
This gauge invariance allows us to eliminate one  superfield, though the formulas become more cumbersome when written in terms of the $N-1$ independent gauge-invariant combinations. 
Instead we will keep working
in the redundant description in terms of $N$ superfields, keeping  gauge invariance under the shift symmetry  (\ref{shiftsf}) manifest at all stages.

\subsubsection{Component form}
Let us now discuss the Coulomb branch mechanics in terms of component fields. Working out the
superspace integrals the relative Lagrangian (\ref{Lsuspace1}-\ref{Lsuspace4}) takes the form\footnote{See Appendix \ref{Appsusy}, (\ref{Lsusy1}) for  the derivation of the component form of
$L^{(2)}$. For the derivation of $L^{(1)}$, which is somewhat involved, we refer to \cite{Ivanov:2002pc}. As a shortcut, it is straightforward to derive the presence of the first term in $L^{(1)}$, from which the remaining terms are fixed by supersymmetry \cite{Denef:2002ru}.}  originally derived in 
\cite{Denef:2002ru}:
\bea
L &=&L^{(0)} + L^{(1)} + L^{(2)}\label{Lag1}\\
L^{(0)} &=&  - f_a  D^a \nonu
L^{(1)} &=&  - U_a  D^a+  A_{ia}  \dot x^{ia} + \pa_{ib}U_{a}  \bar \l^a \s_i \l^b\nonu
L^{(2)} &=& \half G_{ab}  \left(\dot x^{ia} \dot x^{ib} + D^a D^b + i (\bar \l^a \dot \l^b - \dot{\bar \l}^b \l^a)\right)\nonu
&-&  {1 \over 2}  \pa_{ic}  G_{ab}   \left(  \bar \l^a\s_i \l^b   D^c+ \e_{ijk}\bar \l^a \s_j \l^b \dot x^{kc} \right) 
- {1 \over 8}  \pa_{jc}\pa_{jd}G_{ab} 
\l^a \e \l^b \bar \l^c \e \bar \l^d.\label{Lagcomp}
\eea
Here, the symbol $\pa_{ia}$ stands for ${\pa \over \pa x^{ia}}$. To obtain this form of the action we have taken the boundary term in (\ref{Lsuspace1}) to be
\be 
L_{\rm bdy } = {1\over 4} \ddot \calh (x).\label{Lbdy}
\ee

We recall that the functions $U_a$ in $L^{(1)}$ are given by (see (\ref{Uquiver})),
\be U_a=\sum_{b, b\neq a} \frac{\kappa_{ab}}{2r_{ab}}.\label{Ux}
\ee 
 The term linear in velocities in $L^{(1)} $ signifies the Lorentz coupling of the dyons to the electro-magnetic field sourced by the other centers,  
more precisely the $A_{ia}(x)$ are given by a superposition of Dirac monopole fields  
\be
A_{ia}=- \sum_{b,b\neq {a}} \kappa_{ab}\,A_i^{\mathrm{D}}(\tilde x_{ab}),\qquad A_i^{\mathrm{D}}(x)=\frac{\epsilon_{ijk} n^j x^k}{2 \,r(x^ln^l-r)}.\label{Adefns}
\ee
Here, $n^i$ is an arbitrary constant unit vector indicating the direction of the Dirac string, and we   defined
\begin{equation}
\tilde x_{ab}=\begin{cases}
x_{ab}=x_a-x_b&\quad\mbox{when } a<b\\
x_{ba}=x_b-x_a&\quad\mbox{when } a>b.
\end{cases}\label{tildex}
\end{equation}

The term $L^{(2)}$ containing two time derivatives describes a supersymmetric nonlinear  sigma model. The  target space metric $G_{ab}$ is a partial trace of the Hessian matrix of the potential $\calh$ defined in (\ref{Hnonscaling}),
\be \label{Hessepot}
G_{ab} =\half\pa_{ia}\pa_{ib}  \calh.
\ee
We will therefore  refer to $\calh$ as the `Hesse potential' in what follows. As we already remarked above, $\calh$ is not unique and can be redefined by `Hessian transformations', such as (\ref{Hesstransf}), by adding a function with vanishing   Hessian matrix.

In addition to experiencing a magnetic field and a nontrivial geometry, the D-brane centers also move in a potential, which can be obtained from integrating out the auxilary fields $D^a$. This is a little bit subtle, since as we pointed out in (\ref{shiftsf}) we are working in a redundant description  which is gauge-invariant under overall shifts  of the fields:
\bea
\d_{\rm shift} x^a_i &=& v^a \e_i (t)\label{U1Xs}\nonu
\d_{\rm shift} \l^a &=& v^a \h (t), \qquad  \d_{\rm shift} \bar \l^a = v^a \bar \h (t)\label{U1shift}\\
\d_{\rm shift} D^ a &=& v^a \d (t).\nonumber
\eea
(recall that  $v^a \equiv (1,1,\ldots, 1)$). As a consequence of (\ref{shiftss}), the target space `metric' $G_{ab}$ is actually degenerate and $v^a$ is a null vector:
\be 
G_{ab} v^b =0.
\ee
However in  the models of interest  it is invertible when restricted to   the $N-1$-dimensional space of independent relative coordinates which are gauge-invariant under (\ref{U1Xs}). The resulting inverse is called the 
`projective inverse'  $G^{ab}$ and can be constructed concretely as follows. 
We pick an arbitrary vector $v_a$ satisfying 
\be v_a v^a =1 \label{vnorm} \ee 
and define $G^{ab}$ by\footnote{The pair $G^{ab}$ and $v_a$ form a Euclidean version of the Newton-Cartan geometry appearing in non-relativistic theories of gravity, see e.g. \cite{kunzle1972}.} 
\begin{equation}
v_a v^b+G_{ac}G^{cb}=\delta_a^b,\qquad v_av_bG^{ab}=0.\label{invdef}
\end{equation}
Note that this implies  \be G^{ab} v_b=0 .\ee  The projective inverse $G^{ab}$ can for practical purposes be used as the inverse of $G_{ab}$.
For example, to integrate out the auxiliary fields we write their equation of motion with the help of (\ref{invdef}) as
\be 
D^a = G^{ab} \left( f_b+U_b + \half \pa_{ib} G_{cd}\bar \l^c \s_i \l^d \right) + v_b D^b v^a.\label{eomD}
\ee
The last term can be removed by a gauge transformation (\ref{U1shift}) and drops out when substituting (\ref{eomD}) into the action.
Doing this one finds the potential for the  bosonic coordinates,
\be
V = \frac{1}{2} G^{ab}(f_a+U_a)(f_b+U_b).
\ee
Similarly, one can check that this expression does not depend on the   choice of $v_a$ satisfying (\ref{vnorm}). 

Let us now discuss the symmetries of the Lagrangian in component form. From the component expansion (\ref{Phicomp}) of the superfield we find that a general symmetry transformation acts on the fields as 
\bea 
\d &\equiv&  u \d_H  + \x_\a \d_{Q^\a} +\bar  \x^\a \d_{\bar Q_\a} + r^i \d_{R^i} +\tilde  r^i \d_{\tilde R^i} \nonumber\\
\d x^{ia} &=& u \dot x^{ia}+ i \bar \l^a \s^i \x - i \bar \x\s^i \l^a +  \e_{ijk} r^j x^{ka}\nonumber\\
\d \l^a &=& u \dot\l^a + \dot x^{ia} \s_i \x + i D^a \x - {i \over 2} r^i \s_i \l^a + {i \over 2} \tilde r^3 \l^a + \tilde r^- (\e \bar \l^a)_\a \label{comptransf}\\
\d \bar \l^a  &=& u \dot{\bar \l}^a + \dot x^{ia} \bar \x \s_i - i D^a \bar \x+ {i \over 2} r^i \bar \l^ a\s_i  - {i \over 2}\tilde r^3 \bar  \l^a + \tilde r^+  (\e \l^a)^\a \nonumber\\
\d D^a &=& u \dot D^a - \dot{\bar \l}^a\x - \bar \x \dot \l^a .\nonumber
\eea
A general action of the form (\ref{Lag1}-\ref{Lagcomp}) is invariant under $\caln = 4$ supersymmetry transformations 
(with parameter $\x, \bar \x$) provided the couplings satisfy (\ref{Hessepot}) and\footnote{While it  is not  necessary for supersymmetry to impose the last constraint (\ref{symmK}) on $\calh$, it is satisfied in our model since our $\calh$ is a sum of pairwise terms depending only on the coordinate differences, see
	(\ref{Hnonscaling}). If $\calh$ is instead an arbitrary function, the Lagrangian contains additional terms, which are given for completeness in (\ref{Lsusy2}).  }
\bea
\pa_{ia} U_{b} &=& \pa_{ib} U_{a}  =\e_{ijk} \pa_{jb} A_{ka},
 \label{UAids} \\
\pa_{ia}\pa_{jb} \calh &=& \pa_{ib}\pa_{ja} \calh .\label{symmK}
\eea
These restrictions on the couplings imposed by $\caln=4$  supersymmetry can, as usual, be interpreted in terms  of  geometric structures on the target space of the sigma model. These structures are easier to analyze and relate to the literature in a reformulation of the model, as  we will discuss in section \ref{HKTsec}.

Invariance under the $su(2)$ R-symmetries is due to to the fact that $U_a$ and  $G_{ab}$ are rotationally invariant functions while the gauge potential $A_{ia}$ transforms as a vector up to a gauge transformation,
\bea
\e_{ijk} \pa_{ja} U_{b} x^{ka} &=& \e_{ijk}  \pa_{ja}G_{bc} x^{ka}=0, \label{r1} \\
 \e_{ikl} \pa_{kb} A_{ja}   x^{bl} 
&=&  \e_{ijk} A_{ka} +\pa_{ja} M_{i},\label{deltaRA}
\eea 
for some functions $M_i (x)$. Their explicit form, which is derived in Appendix \ref{rotap} and will be needed below, is 
\be 
M_i = \e_{ijk} A_{aj} x^a_k + U_a x^a_i.\label{Mexpr}
\ee
The $su(2)$ R-symmetry has the interpretation of the angular momentum of the 3+1 dimensional
D-brane system. It can be shown  (e.g. using (\ref{Rcharges})   below), that  on classical ground states satisfying $\dot x_{ia}= \l^a = \bar \l^a =0$, the $su(2)$ Noether charges reduce to the term $ U_a x^a_i$
in (\ref{Mexpr}).  This can be rewritten as
\be 
\left.R_i\right|_{\rm vac} =  U_a x^a_i =\half \sum_{a,b, a<b} {\k_{ab} \over r_{ab}} x^{ab}_i,
\ee
which is precisely  the expression for the ADM angular momentum of the corresponding 3+1 dimensional multi-centered supergravity solution \cite{Denef:2000nb}.

\section{Superconformal invariance in the AdS$_2$ scaling limit}\label{confsec}
In this section we demonstrate the emergence of a $D(2,1;0)$  superconformal symmetry when the DSZ parameters $\kappa_{ab}$ allow for a  scaling limit in which the supergravity solution develops a deep AdS$_2$ throat. 
We  check superconformal invariance in superspace  and explicitly compute the Noether charges in the component formalism. 
Conformal invariance of the three-centered case was analyzed in detail in \cite{Anninos:2013nra}, while $D(2,1;0)$ invariance of a single-particle cousin of our models\footnote{See also \cite{Ivanov:1988it}, \cite{deAzcarraga:1998ni}, \cite{Maloney:1999dv}, \cite{Donets:2000sx}, \cite{Kuznetsova:2011je} for studies of related models with $\caln=4$ superconformal symmetry.} was established in \cite{Ivanov:2002pc}. 

\subsection{Scaling charges and AdS$_2$ limit}
It is known from the supergravity description that for certain charges the dyonic centers can be placed arbitrarily close together, see e.g. \cite{Denef:2007vg}. As the centers approach each other in coordinate space a diverging gravitational warping keeps them at finite physical distance. This regime can be explored by an appropriate limit of the gravitational solution leading to gravitational multi-center configurations  with AdS$_2\times$S$^2$ asymptotics \cite{Mirfendereski:2018tob}, see also \cite{Bena:2018bbd}. Complementarily it was shown in \cite{Anninos:2013nra} that the same limit of the quiver description leads to the emergence of conformal symmetry. From an operational point of view the limit amounts to redefining the variables as\footnote{The supergravity metric takes the form $ds^2=-\Sigma(x^i,x^{ia},f_a)^{-1}(dt+\omega)^2+\Sigma(x^i,x^{ia},f_a)dx^idx^i$, where $(t,x^i)$ are the space-time coordinates and $x^{ia}$ are the positions of the dyonic centers. The homogeneity property $\Sigma(s x^i, sx^{ia}, f_a)=s^{-2}\Sigma (x^i,x^{ia},sf_a)$ and the form of the metric ensure that the rescaling $t\rightarrow s^{-1} t, x^i\rightarrow s x, x^{ia}\rightarrow s x^{ia}, f_a\rightarrow f_a$, as in \eqref{resc}, are equivalent to the rescaling $t\rightarrow t, x^i\rightarrow x, x^{ia}\rightarrow x^{ia}, f_a\rightarrow s f_a$ which is the form used in the near limit as defined in \cite{Mirfendereski:2018tob}.} 
\begin{equation}
t\rightarrow s^{-1} t\,,\quad x^{ia}\rightarrow s x^{ia}\,,\quad \lambda^a_\alpha\rightarrow s^{3/2} \lambda^a_\alpha\,,\quad D^a\rightarrow s^2 D^a\label{resc}
\end{equation}
and then taking the limit where $s$ goes to zero. In this limit the action remains finite and has as only effect that the relative masses $\mu_{ab}$ and FI couplings $f_a$ are made to vanish. In particular the form of the Lagrangian (\ref{Lag1}-\ref{Lagcomp}) remains intact, except that some of the couplings  simplify, 
\begin{align}
f_a =& 0, &
\calh =& -\sum_{a,b, b\neq a} \frac{|\kappa_{ab}|}{4r_{ab}}\log r_{ab}, & 
G_{ab}=& \delta_{ab}\left(\sum_{c,c\neq {a}} \frac{|\kappa_{ac}|}{4r_{ac}^3}\right)-\frac{|\kappa_{ab}|}{4r_{ab}^3}. \label{scalingcouplings}
\end{align}
The scaling limit does not change the term\footnote{In fact, it can be shown  \cite{Ivanov:2002pc} that $L^{(1)}$ is invariant under $D(2,1;\a )$ for any $\a$.} $L^{(1)}$  in the Lagrangian (see (\ref{Ux},\ref{Adefns})).
Since these new expressions keep on satisfying \eqref{Hessepot} and \eqref{UAids} it follows that the limit preserves supersymmetry, as is of course manifest from the superspace point of view. As we will now discuss it additionally enhances it to a superconformal symmetry.

\subsection{Superconformal invariance}
\subsubsection{Superspace}
The action of the $\caln = 4$ superalgebra (\ref{Poincsusy}, \ref{Rsymm}) on $\mathbb{R}^{1|4}$ superspace can be extended to an action of the superconformal algebra $D(2,1;\a)$  (see e.g. \cite{Okazaki:2015pfa} for a review). Here, $\a$ is a continuous parameter and as we will presently see\footnote{A quick way to determine $\a$ is by using (see e.g. \cite{Kuznetsova:2011je})  that the scale weight of the superfield $\F_i$ is $-\a$, and for $L^{(2)}$ to be invariant we need this weight to be one.}, the scaling limit of the quiver mechanics is invariant under $D(2,1;-1)$, which is isomorphic to $D(2,1;0)$ upon exchanging the role of the two $su(2)$ R-symmetries. 
This algebra contains two additional  bosonic generators $D$ and $K$ which generate dilatations and special conformal transformations respectively, and four additional fermionic superconformal generators  $S^\a$ and $\overline S_\a$. These are given explicitly by (see \cite{Ivanov:2002pc} with $\a = -1$):
\bea 
D &=& t \pa_t + \half \left( \theta \pa_ \theta  + \bar  \theta \pa_{\bar  \theta}\right)\\
K &=& \left( t^2 - (\bar \theta \theta)^2\right) \pa_t + (t + i \bar \theta \theta)  \theta \pa_\theta  + (t - i \bar \theta \theta) \bar \theta \pa_{\bar \theta} \\
S^\a &=&   t Q^\a + i \bar \theta \theta D^\a + 2 i \bar \theta^\a  \bar \theta \pa_{\bar \theta}\\
\overline S_\a &=&    t \overline Q_\a - i \bar \theta \theta \overline D_\a + 2 i  \theta_\a   \theta \pa_{ \theta}.
\eea
The new nonvanishing  (anti-)commutators  in addition to (\ref{Poincsusy}, \ref{Rsymm}) are
\begin{align}
[H,D] =& H, & [H,K] =& 2 D, & [D,K] =& K,\\
[D,Q^\a] =& -\half  Q^\a , & [D,\overline Q_\a] =& - \half \overline Q_\a,& &\\
[K,Q^\a] =& -S^\a , & [K,\overline Q_\a] =&- \overline S_\a,& &\\
[H,S^\a] =&  Q^\a , & [H,\overline S_\a] =&  \overline Q_\a,& &\\
[D,S^\a] =& \half S^\a , & [D,\overline S_\a] =& \half \overline S_\a,& &\\
\{ Q^\a, \overline S_\b \} = &  2i D \d^\a_\b - 2 R_i \s_{i\a}^{\ \ \b} , & \{\overline Q_\a,  S^\b \} = &  2i D \d^\a_\b + 2 R_i \s_{i\a}^{\ \ \b},  &
 \{ S^\a, \overline S_\b \} =& 2 i \d^\a_\b K, \\
[ R_i , S^\a ] =&- {i \over 2} ( S\s_i)^\a, & [ R_i ,\overline S_\a ] =& {i \over 2} (\s_i \overline S )_\a,\\
[\tilde R_+ , S^\a ] =& (\e \overline S)^\a, & [\tilde R_- , \overline S^\a ] =&- (\e  S)_\a,\\
[\tilde R_3 , S^\a ] =&{i \over 2} S^\a, & [\tilde R_3 , \overline S_\a ] =&- {i \over 2} \overline S_\a.
\label{suconfKV}
\end{align}
If we disregard the generators $\tilde R_i$ we obtain the subalgebra $psu(1,1|2)$. The  $\tilde R_i$ act on  $psu(1,1|2)$ as outer automorphisms.

The transformation of the fields under the $D(2,1;-1)$ algebra is obtained from the starting assumption that the real superfield $V$ transforms as  a scalar under the full superconformal algebra:
\be 
\d V = (u H + v D + w K + r^i R_i+ \tilde r^i \tilde R_i + \x Q + \bar \x \overline Q + \h S +\bar \h \overline S ) V.
\ee
To find the transformation of the superfield $\F_i = \half D \s_i \overline D V $ we should commute the algebra generators  through the operator $D \s_i \overline D $. Doing this we find
\bea 
\d \F_i &=& \left( \dot P 
- 2i (\bar \theta \dot \S  - \dot{\overline \S} \theta ) \right)\F_i  + \e_{ijk}\left(r^j -  \ddot P \bar \theta \s_j \theta - 2\bar \theta \s_j \dot \S + 2\dot{\overline \S} \s_j \theta\right)  \F_k \nonu
&& +(u H + v D + w K + r^i R_i+ \tilde r^i \tilde R_i + \x Q + \bar \x \overline Q + \h S +\bar \h \overline S ) \F_i ,\label{deltaPhi}
\eea
where we have defined, for later convenience, the following time-dependent combinations of parameters
\be
P \equiv u + vt + w t^2, \qquad
\S_\a \equiv  \x_\a  + \h_\a  t, \qquad \bar \S^\a \equiv \bar \x^\a  + \bar \h^\a  t.
\ee

Using the above relations it is straightforward to check the $D(2,1;-1)$-invariance of the superspace form of the action (\ref{Lsuspace1}). 
It suffices to check invariance under  conformal transformations,
\be 
\d_{\rm conf} \equiv u \d_H + v \d_D + w \d_K,
\ee
since, in combination with the invariance under supersymmetries and $R$-symmetries already demonstrated, this implies invariance under the  full algebra. 
Since the superspace Lagrangian density depends on the  combinations $|\F^{ab}|$, the rotation term in (\ref{deltaPhi}) doesn't contribute and we have
\bea
 \d_{\rm conf}\int d^4 \theta  \call &=& \int d^4 \theta \left( {\pa \call \over \pa \F^{ia}} \F^{ia} \dot P + (u H + v D + w K) \call\right)\nonu
&=& \dot P \int d^4 \theta \left( {\pa \call \over \pa \F^{ia}} \F^{ia}  +  \call   \right)+ \ldots\nonu
&=&  \dot P {d^2 \over dt^2} \left( \sum_{a,b,a \neq b} 
{|\k_{ab}|\over 16 |x^{ab}|}
\right)+ \ldots\eea
where, in the second line, we have partially integrated in superspace and dropped total time derivatives, and, in the last line, we used the explicit form (\ref{Lsuspace1}) and the identity \eqref{int1F}. The result is a total time derivative because $\dot P$ is a linear function of time. 

\subsubsection{Component fields}
From the superspace transformation law (\ref{deltaPhi}) and the component decomposition (\ref{Phicomp}) we can work out the action of $D(2,1;-1)$ on the component fields:
\bea 
\d &\equiv&  u \d_H + v \d_D + w \d_K + \x_\a \d_{Q^\a} +\bar  \x^\a \d_{\bar Q_\a} + \h_\a \d_{S^\a} +\bar  \h^\a \d_{\bar S_\a} + r^i \d_{R^i} +\tilde r^i \d_{\tilde R^i}\nonu
\d x^{ia} &=&   \dot P  x^{ia} + P \dot x^{ia} +  \e_{ijk} r^j x^{ka} +  i \bar \l^a \s^i \S - i \bar \S\s^i \l^a  \nonu
\d\l^a &=& {3 \over 2}\dot P  \l^a + P \dot\l^a   + \dot x^{ia} \s_i \S + i D^a \S + 2 x^{ia} \s_i \dot \S - {i \over 2} r^i \s_i \l^a  + {i \over 2} \tilde r^3 \l^a + \tilde r^- (\e \bar \l^a)_\a\nonu 
 \d \bar \l^{ a} &=&{3 \over 2}\dot P \bar \l^a + P \dot{\bar \l}^a + \dot x^{ia} \bar \S \s_i - i D^a \bar \S^\a + 2 x^{ia}\dot{ \bar \S} \s_i+{i \over 2} r^i \bar \l^ a\s_i  
  - {i \over 2}\tilde r^3 \bar  \l^a + \tilde r^+  (\e \l^a)^\a \nonu
\d D^a &=& 2 \dot P  D^a + P \dot D^a  - \dot{\bar \l}^a\S  - \bar \S \dot \l^a - 3 \bar \l^a \dot \S -  3\dot{\bar \S} \l^a .\label{comptrans}
\eea
In particular, under the subalgebra of conformal transformations, the fields $x^{ia}, \l^a_\a$ and $D^a$ transform as primary fields of weight $\Delta=1, {3 \over 2},$ and 2, respectively (see Appendix \ref{Appprimary} for the definition of a primary field).

\subsection{Noether charges}
We  now  compute the conserved Noether charges  associated to the symmetry generators.
We recall that,  if the Lagrangian transforms under a symmetry by a total derivative,
\be
\d_{\rm sym} L = {d \over dt} B_{\rm sym}, 
\ee
the associated conserved charge is given by
\be 
Q_{\rm sym} =  {\pa L \over \pa \dot x^a_i} \d_{\rm sym} x^a_i +   {\pa_R L \over \pa \dot \l^a_\a} \d_{\rm sym} \l^a_\a+  {\pa_R L \over \pa \dot{\bar \l}^{\a a}}  \d_{\rm sym} \bar \l^{\a a}- B_{\rm sym}.\label{chargeformula}
\ee
In writing this formula we have taken the convention that derivatives with respect to the fermionic fields $\l^a, \bar \l^a,$ act from the right\footnote{In contrast, following standard conventions, derivatives with respect to the superspace coordinates $\theta_\a, \bar \theta^\a$ in earlier sections were defined to act from the left.}, as indicated by the subscript $_R$, 
see Appendix \ref{fermder} for more details.

While the relevant boundary terms $B_{\rm sym}$ can also be derived from the superspace form of the action, the analysis is less cumbersome in the component formalism. 
We will discuss in turn the Noether charges for conformal, R- and fermionic symmetries.  
\subsubsection{Conformal}
The component action is invariant under conformal transformations thanks to the properties
\bea
x^{ic}\partial_{ic}G_{ab}&=&-3G_{ab}, \qquad G_{ab} x^{ib} = - \pa_{ia} \left( G_{ab}x^{ja}x^{jb} \right),\label{hompropG} \\
x^{ib}\partial_{ib}U_a&=&-U_a, \qquad x^{jb}\partial_{jb}A_{ia}=x^{jb}\partial_{ia}A_{jb}.
\eea
These are a specific case of the general requirements on the target space geometry derived in \cite{Papadopoulos:2000ka}: the metric possesses a conformal Killing vector (in our case $x^{ia} \pa_{ia}$), whose associated one-form is exact. In addition, the potential should scale with the proper weight and  the gauge connection should be invariant up to a gauge transformation. Making use of the additional algebraic identity (\ref{id1}) satisfied by  $A_{ia}$,
one finds that the Lagrangian transforms as
\begin{equation}
\delta_{\rm conf}  L=\frac{d}{dt}\left(PL - \ddot P G_{ab}x^{ia}x^{ib} \right).
\end{equation}
This leads to the conformal Noether charges\footnote{We use the same symbol for the symmetry generator and the associated Noether charge in  our particular models, hopefully without causing confusion.}
\bea
H&=&  U_a D^a-\pa_{ia} U_b\bar \lambda^a\sigma_i\lambda^b +\frac{1}{2}G_{ab}\left(\dot x^{ia}\dot x^{ib}-D^a D^b\right)+\frac{1}{2}\partial_{ic}G_{ab}\bar \lambda^a\sigma_i\lambda^b D^c\nonu &&+\frac{1}{8}\partial_{jc}\partial_{jd}G_{ab}\lambda^a\epsilon\lambda^b\bar\lambda^c\epsilon\bar \lambda^d\label{HNoether}\\
D&=&t H+G_{ab}x^{ia}\dot x^{ib}\nonu
K&=&t^2 H+2tG_{ab}x^{ia}\dot x^{ib}+2G_{ab}x^{ia}x^{ib}\label{DKNoether}
\eea
We note that dilatations and special conformal transformations are symmetries of the action which however do not commute with time translations;  their Noether charges depend explicitly on time.
These charges are however conserved on-shell, $\dot D \approx 0, \dot K \approx 0$, due to the on-shell identity
\be
{d \over dt} \left(G_{ab} x^a_i \dot x^b_i \right) \approx - H
\ee
as well as the algebraic identity (\ref{id2}).

\subsubsection{R-symmetry}
Next we compute the Noether charges for the $su(2)$ R-symmetry
under which the  $x_i^a$ transform as triplets and the spinors $\l^a_\a, \bar \l^{\a a}$ 
as doublets (\ref{Rsymm}). Most terms in the Lagrangian are  invariant under this symmetry, except for the magnetic coupling $ A_{ia} \dot x^{ia}$  which, as follows from (\ref{deltaRA}), transforms  by  a total derivative
\be 
\d_{R^i}(A_{ja} \dot x^a_j) =- {d M_i \over dt},
\ee
where the functions $M_i$ were given in (\ref{Mexpr}). The $su(2)$ Noether charges $R_i$ therefore take the form 
\be
R_i =  {\pa L \over \pa \dot x^a_j}\d_{R^i} x^a_j +   {\pa_R L \over \pa \dot \l^a_\a} \d_{R^i} \l^a_\a+  {\pa_R L \over \pa \dot{\bar \l}^{\a a}} \d_{R^i} \bar \l^{\a a}+ M_i .\label{Rcharges}
\ee

As for the second R-symmetry $\widetilde{ su(2)}$, one can check that the Lagrangian $L$ is invariant and therefore the Noether charges are of the form 
\be
\tilde R_i =    {\pa_R L \over \pa \dot \l^a_\a} \d_{\tilde R^i} \l^a_\a+  {\pa_R L \over \pa \dot{\bar \l}^{\a a}} \d_{\tilde R^i} \bar \l^{\a a}. \label{Rtcharges}
\ee

\subsubsection{Fermionic charges}
We now turn to the derivation of the Noether charges for the fermionic symmetries, i.e. the four Poincar\'e supercharges and four conformal supercharges.
To simplify the computation, we are free to add a total time derivative to the Lagrangian. This does not influence the equations of motion and one easily sees that the expression (\ref{chargeformula}) for the Noether charge does not change under such an addition. 
We define $L_+$ ($L_-$) to be the partially integrated Lagrangians in which $\bar \l^a$ ($\l^a$) do not appear with time derivatives:
\be 
L_\pm =  L \pm{d\over dt }\left(  {i\over 2}  G_{ab} \bar \l^a \l^b \right).\label{Lpm}
	\ee
Let us first discuss invariance  under $\d_{Q^\a}$ and $\d_{S^\a}$, which we combine into
\be
\d_\S \equiv \x_\a \d_{Q^\a} + \h_\a \d_{S^\a} .
\ee
For these transformations, it is convenient to work with $L_-$:  since  ${\pa L_- 
	\over \pa \dot \l^a} =0$ and also $\d_\S\bar \l^a =0$, the second and third terms on the RHS of (\ref{chargeformula}) vanish, while also the boundary term (the last term on the RHS of  (\ref{chargeformula}))  becomes simpler. 
One finds in particular that $L_-$ is invariant up to the boundary term 
\be 
\d_\S L_- = {d \over dt} \left( U_a \l^a \S + i A_{ia} \bar \l^a \s^i \S - 2 i G_{ab}x^{ia} \bar \l^b \s_i \dot \S \right).
\ee
The Noether charges are then
\bea 
Q^\a &=&- U_a \bar \l^{a\a} + i \left({\pa L_- \over \pa \dot x^{ia} }- A_{ia} \right) (\bar \l^a \s^i)^\a, \nonu
S^\a &=& t Q^\a + 2i G_{ab}x^{ia}  (\bar \l^b \s^i)^\a .\label{QS}
\eea 
To check the conservation of the conformal supercharges, $\dot S^\a \approx 0$,  one can show that the following identity holds on-shell:
\be
Q^\a \approx - 2 i {d\over dt} \left( G_{ab} x^a_i (\bar \l^b \s_i)^\a \right).\label{Qisder}
\ee
This can be shown to hold using the identities (\ref{hompropG}, \ref{id4}, \ref{id5}).

Similarly, to address the  symmetry under $\d_{\overline  Q_\a}$ and $\d_{\overline  S_\a}$ it is convenient to work with the Lagrangian $L_+$. One finds in this way the Noether charges
\bea 
\overline Q_\a &=&- U_a  \l^{a}_\a-  i \left({\pa L_+ \over \pa \dot x^{ia} }- A_{ia} \right) (\s^i \l^a )_\a ,\nonu
\overline S_\a &=& t \overline Q_\a - 2i G_{ab}x^{ia}  (\s^i \l^b )_\a. \label{QbSb}
\eea 
These are the complex conjugate expressions of (\ref{QS}) as expected.

\subsection{Canonical variables  and Poisson bracket algebra}\label{Hamsec}
In this subsection we describe the superconformal Coulomb branch mechanics  in terms of canonical variabes in the  Hamiltonian formalism.
 This  paves the way for quantization of the system, which will be considered elsewhere.
 As a check we  also explicitly compute the Poisson bracket algebra of the conserved charges
 and find that this yields the $D(2,1;-1) \simeq D(2,1;0)$ algebra.
 
 The canonical analysis is somewhat simpler if we add a total derivative to the Lagrangian (\ref{Lagcomp}) and work with $L_+$ or $L_-$ introduced in  (\ref{Lpm}) so that either $\bar \l^a$ or $ \l^a$
 appears without time derivative. We will choose the former option and work with 
 $L_+$ which we spell out here:
\bea
 L_+ &=& L^{(1)} + L^{(2)}_+\label{Lag}\nonu
L^{(1)} &=&  - U_a  D^a+  A_{ia}  \dot x^{ia} + \pa_{ib}U_{a}  \bar \l^a \s_i \l^b \label{L+comp}\\
L^{(2)}_+ &=& 
\half G_{ab}  \left(\dot x^a_i \dot x^b_i + D^a D^b \right) +  i \bar \l_a \dot \l^b \nonu
&-&  {1 \over 2}  \pa_{ic}  G_{ab}   \left(  \bar \l^a\s_i \l^b   D^c- i \dot x^{ic} \bar \l^a \l^b+ \e_{ijk}\bar \l^a \s_j \l^b \dot x^{kc} \right) 
- {1 \over 8}  \pa_{jc}\pa_{jd}G_{ab} 
\l^a \e \l^b \bar \l^c \e \bar \l^d,\nonumber
 \eea
 where we have defined fermionic fields with index lowered as
 \be
 \bar \l_a \equiv G_{ab} \bar \l^b.
 \ee
 The relevant coupling functions were given in (\ref{Ux}, \ref{Adefns}, \ref{scalingcouplings}). 
  
  We find for the bosonic canonical momenta
\be 
p_{ ai}\equiv{\pa L \over\pa \dot x^a_i} =
A_{ia} + G_{ab}\dot x^a_i+ {i \over 2} {\pa_{ia} G_{bc} } \bar \l^b \l^c -\half \e_{ijk} {\pa_{ja} G_{bc}} \bar \l^b \s_k \l^c.\label{pbos}
\ee
We note from this expression that  the  $p_{ai}$ have an imaginary part 
\be 
p_{ai}^* = p_{ai} - i  {\pa_{ia} G_{bc}} \bar \l^b \l^c.\label{realityp}
\ee
This originates from the fact that $L_+$, in contrast to the original Lagrangian $L$ in (\ref{Lagcomp}), has an imaginary part which is a total derivative.

For the canonical formulation of the fermionic sector, it is easiest\footnote{Alternatively, one can regard both $\l^a$ and  $\bar \l_a$ as configuration space variables. Then there are two second class constraints, $0 = \p_a - i \bar \l_a = \bar \p^a$, and the resulting Dirac bracket again leads   to (\ref{elPB}).} to note, as in \cite{Faddeev:1988qp}, that the action is already in first order form with  the role of the momenta conjugate to coordinates $\l^a_\a$ played by
\be 
\p_a^\a \equiv i \bar \l_a^\a.\label{piferm}
\ee
The Poisson brackets for the fermions  are $\{ \bar \l^{a\a} ,\bar \p_{b\b}\} = \d^a_b \d_\a^\b$, and we are led to the nonvanishing brackets\footnote{In computing Poisson brackets of phase space quantities it is important to keep in mind that the bosonic momenta $p_{ia}$ Poisson-commute with $\l^a$ and $\bar \l_a$ but not with $\l_a,\bar \l^a$.} for the full theory
\be 
\{ x^{ia} , p_{jb} \} = \d^a_b \d_{ij}, \qquad \{ \l^a_\a , \bar \l_{b}^\b\} = -i\d^a_b \d_\a^\b.\label{elPB}
\ee

 We should remark that, strictly speaking, there are also constraints involving the auxiliary fields $D^a$
 \be 
D^a= G^{ab} \left( U_b + \half { \pa_{ib} G_{cd} }\bar \l^c \s_i \l^d \right), \qquad  p_{D^a} = 0. \label{Dsol}
 \ee
 These  are, e.g. using the Dirac formalism, trivially taken care of by substituting (\ref{Dsol}) everywhere, in particular  $D^a $ has nontrivial Dirac brackets with $x, p, \l, \bar \l$.


The canonical Hamiltonian agrees with the time translation Noether charge (\ref{HNoether}) and takes the form
\bea
H &=& p_{ia} \dot x^{ia}+ \p_a \dot \l^a - L \nonu
&=& \half P_{ia} G^{ab} P_{ib} + \half G_{ab} D^a D^b - C_{iab} \bar \l^a \s_i \l^b+ {1\over 8} \pa_{icid} G_{ab} \l^a \e \l^b \bar \l^c \e \bar \l^d,
\eea
where we defined the `kinetic momentum' $P_{ai}$ as
\be 
P_{ai}   = G_{ab} \dot x^b_i =p_{ai} - A_{ia}- {i \over 2} {\pa_{ia} G_{bc} } \bar \l^b \l^c +\half \e_{ijk} {\pa_{ja} G_{bc}} \bar \l^b \s_k \l^c.
\ee

Our expressions for the remaining Noether charges (\ref{DKNoether}, \ref{Rcharges}, \ref{Rtcharges}, \ref{QS}, \ref{QbSb}) can be written\footnote{The expressions for $D,K, R_i$ were simplified using the identities (\ref{hompropG}) and (\ref{id3}).} in phase-space form:
\bea
D &=& t H + x^a_i p_{ai} + {3 i \over 2}  \bar \l_a \l^a\nonu
K&=& t^2 H + 2t  x^a_i  p_{ai}+ 3i t   \bar \l_a \l^a+ 2G_{ab}x^{ia}x^{ib}\nonu
Q^\a &=& -U_a\bar \l^{a\a} +i \left(p_{ia} - A_{ia}- i \pa_{ia}G_{bc} \bar \l^b \l^c\right) (\bar \l^a \s_i)^\a\nonu
\overline {Q}_\a &=& -U_a\l^a_\a - i \left(p_{ia} - A_{ia} \right) (\s_i \l^a )_\a \nonu
S^\a &=&  t Q^\a + 2 i x^a_i (\bar \l_a \s_i)^\a,\qquad
\overline S_\a =  t \overline Q_\a - 2 i G_{ab} x^a_i ( \s_i \l^b)_\a\label{Ss}\nonu
R_i 
&=& \e_{ijk}x^a_j ( p_{ka}- A_{ka})+ U_a x^a_i +\half  \bar \l_a \s_i  \l^a \nonu
\tilde R_3 &=& - \half \bar \l^a \l_a,\qquad
\tilde R_+ = {i \over 2} \l^a \e \l_a, \qquad \tilde R_- = {i \over 2} \bar \l_a \e \bar \l^a.
\label{genrspq} 
\eea
One checks  that the  $\overline Q_\a$ are the complex conjugates of   $Q^\a$ using the non-reality of $p_{ai}$, see (\ref{realityp}).
As a check on these expressions, we have verified that the Noether charges above generate the symmetry transformations of the fields (\ref{comptrans}) through Poisson brackets in the sense that
\be 
\{ Q_{\rm sym} , \psi \}= - \d_{\rm sym} \psi.
\ee

Before continuing we should remark on the gauge invariance (\ref{U1shift}) acting as overall shifts on the fields. These correspond to first class constraints
\bea 
p_{ai} v^a =0, \qquad   \p_a v^a =0
\eea
as can be seen from (\ref{pbos},\ref{piferm}). The symmetry generators given above are gauge-invariant observables, indeed one easily checks that they weakly Poisson-commute with the constraints
\be 
\{ Q_{\rm sym}, p_{ai} v^a \} \approx \{ Q_{\rm sym}, \p_{a} v^a \} \approx 0.
\ee 
To deal with these constraints we could follow the standard procedure of fixing  the gauge, e.g. by eliminating one of the canonical coordinates, working out the Dirac bracket on the reduced phase space and then quantize this bracket. As already mentioned, this leads to more cumbersome and less symmetric formulas. Here we will rather keep working with the extended phase space (\ref{elPB}) and  leave imposing the constraints until after quantization, where on the Hilbert space they will take the form
\be 
\hat p_{ai} v^a | \psi \rangle  =0, \qquad  \hat  \p_a v^a  | \psi \rangle =0.\label{constrHilb}
\ee
This is the so-called Dirac quantization, and for the simple gauge invariance considered here the two approaches can be shown to be completely equivalent (see \cite{Henneaux:1992ig}, ch. 13).  Imposing the constraints (\ref{constrHilb}) is  straightforward, for example the first constraint is satisfied if we  take  the wavefunction to  depend only on the relative coordinates.

We are now ready to compute the Poisson brackets of the generators (\ref{genrspq}). Our sign conventions   for computing brackets
involving Grassmann-valued  fields are spelled out in appendix \ref{fermder}.
A shortcut to obtaining the Poisson brackets with the Hamiltonian is to use the conservation law
for the generators which depend explicitly on time: 
\be
0 = {d Q_{\rm sym} \over dt} =  {\pa Q_{\rm sym} \over \pa t} + \{Q_{\rm sym}, H\},
\ee 
from which we find
\begin{align}
\{H,D\} =& H, & \{H,K\} =& 2 D,\\
\{H, S^\a\} =& Q^\a, & \{H,\overline S_\a\} =& \overline Q^\a.
\end{align}
The remaining Poisson brackets can be worked out using the computational rules (\ref{PBrules}) and reality properties (\ref{PBreality}). We find\footnote{To derive the brackets (\ref{PBrot}) and (\ref{PBQbS}) one needs the identities (\ref{intGs}) and (\ref{id6}), respectively.}
\begin{align}
 \{ K, Q^\a \} =& - S^\a, & \{ K,\overline Q_\a \} =& - \overline S_\a ,\\
 \{ D, Q^\a \} =& -\frac{1}{2} Q^\a, & \{ D,\overline Q_\a \} =& -\frac{1}{2} \overline Q_\a,\\
 \{ D, S^\a \} =& \frac{1}{2} S^\a, & \{ D,\overline S_\a \} =& \frac{1}{2} \overline S_\a,\\
 \{ R_i , R_j \} =& \e_{ijk}R_k,\label{PBrot}\\
  \{  \tilde{R}_+ , \tilde{R}_- \} =& 2i\tilde{R}_3, &  \{  \tilde{R}_3 , \tilde{R}_\pm \} =&\mp i\tilde{R}_\pm \\
  \{  Q^\a , \overline Q_\b \} =& - 2 i H \d_\b^\a,&
  \{  S^\a , \overline S_\b \} =& - 2 i K \d_\b^\a,\label{odd1}\\
 \{ \overline Q_\a ,S^\b \} =& - 2 i D \d_\a^\b - 2 R_i \s_{i\a}^{\ \ \b}, &\{  Q^\a ,\overline S_\b \} =& -\overline{\{ \overline Q_\a ,S^\b \} }= - 2 i D \d^\a_\b + 2 R_i \s_{i\b}^{\ \ \a},\label{PBQbS}\\
  \{ Q^\a, R_i \} =& \frac{i}{2} \s^{~\a}_{i\b} Q^\b, & \{ \overline Q_\a, R_i \} =&  -\frac{i}{2} \overline Q_\b \s^{~\b}_{i\a},\\
 \{ R_i ,S^\a \} =& - \frac{i}{2} \s_{i\b}^{\ \ \a}S^{\b}, &\{  R_{i} ,\overline S_\a \} =&  \frac{i}{2} \overline{S}_\b\s_{i\a}^{\ \ \b},\\
  \{ \tilde{R}_3, Q^\a \} =& \frac{i}{2} Q^\a, & \{ \tilde{R}_3, \overline Q_\a\} =&  -\frac{i}{2} \overline Q_\a,\\
 \{ \tilde{R}_+, Q^\a \} =& (\overline{Q}\e)^\a, & \{ \tilde{R}_-, \overline Q_\a\} =& \overline{\{ \tilde{R}_+, Q^\a \}}= -(\e Q)_{\a},\\
 \{ \tilde{R}_3, S^\a \} =& \frac{i}{2} S^\a, & \{ \tilde{R}_3, \overline S_\a\} =& -\frac{i}{2} \overline S_\a,\\
 \{ \tilde{R}_+, S^\a \} =& (\overline{S}\e)^\a, & \{ \tilde{R}_-, \overline S_\a\} =& \overline{\{ \tilde{R}_+, S^\a \}}= -(\e S)_{\a}.
\end{align}
As usual (see e.g. \cite{Wess:1992cp} Ch. IV), there are `active vs. passive' sign differences between the  Poisson bracket algebra and the algebra Killing vectors in  (\ref{Poincsusy}, \ref{suconfKV}), which in our conventions show up in the brackets of the odd generators, i.e. (\ref{odd1}, \ref{PBQbS}).

\section{HKT formulation}\label{HKTsec}
The various supermultiplets of ${\cal N}=4$ quantum mechanics are closely related \cite{Ivanov:2003nk, Ivanov:2003tm}. All of them can be connected to the ${\mathbf{(4,4,0)}}$ multiplet \cite{Bellucci:2005xn}, which is the 'simplest', in that it has an equal number of bosons and fermions and is without auxiliary fields. In this section we will reformulate the quiver theory which was presented above in terms of ${\mathbf{(3,4,1)}}$ multiplets -- in terms of ${\mathbf{(4,4,0)}}$ multiplets. Such reformulation has been understood in general from a superspace perspective \cite{Delduc:2006yp}, but we'll take a more pedestrian approach here and present it as a field redefinition in component form. The motivation to consider the ${\mathbf{(4,4,0)}}$ formulation is that the geometry underlying the theory takes a more familiar form, namely that of (weak) hyper-K\"ahler with torsion (HKT). This is especially powerful as a step towards quantization, for which the supersymmetric groundstates can be given a cohomological interpretation in this setting \cite{Smilga:2012wy, Fedoruk:2014jba}.

\subsection{Field redefinitions}\label{fredef}
\begin{figure}
	\centering
	\includegraphics[scale=0.5]{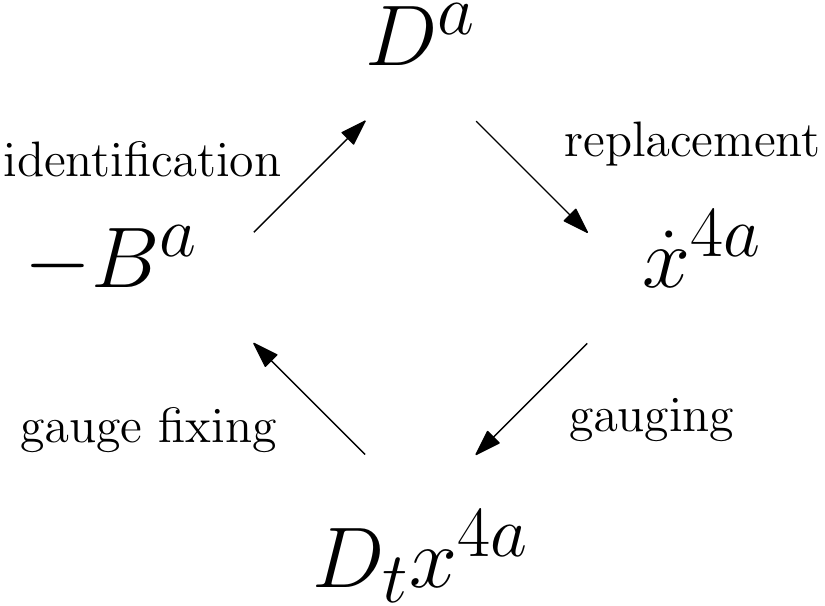}\caption{An operational interpretation of 1d automorphic duality.\label{reffig}}
\end{figure}

We start by introducing $N$ new coordinates $x^{4a}$ such that
\begin{equation}
D^a=\dot x^{4a}.
\end{equation}
Such a replacement of an auxiliary field with the derivative of a new bosonic field is sometimes called '1d automorphic duality' \cite{Delduc:2006yp}. Note that if we reformulate a theory in terms of the $D^a$ in this way the 'new' theory will be invariant under shifts of $x^{4a}$ by construction. This shift symmetry can be gauged by the introduction of a gauge field $B^a$ and the covariant time derivative $D_t x^{4a}=\dot x^{4a}-B^a$. Fixing the gauge so that $x^{4a}$ is constant equates $D_t x^{4a}=-B^a$, and upon identification of $-B^a$ with $D^a$ we are back where we started, see figure \ref{reffig}. Interestingly, and this is special to 1d, the gauge field $B^a$ forms a supermultiplet on its own, without the need of a fermionic partner, so that the bosonic procedure we just sketched is almost trivially supersymmetrized \cite{Delduc:2006yp}. In summary the theory of ${\mathbf{(3,4,1)}}$ multiplets is a gauge fixed form of a theory of ${\mathbf{(4,4,0)}}$ multiplets with a gauged shift symmetry. The $D$-term constraint of the ${\mathbf{(3,4,1)}}$ theory equals the momentum constraint $p_{4a}=0$, which is the 'Gauss constraint' of the gauged ${\mathbf{(4,4,0)}}$ theory.

In this section we will, for simplicity, write the ungauged ${\mathbf{(4,4,0)}}$ theory in terms of $\dot x^{4a}$ with the implicit understanding that to reproduce the results of the previous sections one needs to gauge the shift symmetry in $x^{4a}$ and gaugefix, as in figure \ref{reffig}.

The reformulated theory takes its natural form by collecting the new bosonic coordinates together with the original ones. We thus introduce the 'covariant' notation
\begin{equation}
(x^A)=(x^{\mu a})=(x^{ia},x^{4a})\qquad \mu=1,2,3,4,\ i=1,2,3.
\end{equation}
In parallel we introduce the matrices
\begin{equation}
(\tau^\mu)=(-i\sigma^i,\mathds{1})\qquad (\bar\tau^\mu)=(i\sigma^i,\mathds{1}). \label{taudef}
\end{equation}
Note that the $\tau^i$ generate a quaternion algebra:
\begin{equation}
\tau^i\tau^j=\delta^{ij}\mathds{1}+\epsilon^{ijk}\tau^k.
\end{equation}
In addition there are the following useful relations, that play a role in the derivation of the results presented below:
\begin{align}
(\tau^\mu)_\alpha{}^\beta (\tau^\mu)_\gamma{}^\delta=&(\bar\tau^\mu)_\alpha{}^\beta (\bar\tau^\mu)_\gamma{}^\delta=2\epsilon_{\alpha\gamma}\epsilon^{\beta\delta},& (\tau^\mu)_\alpha{}^\beta (\bar\tau^\mu)_\gamma{}^\delta=&2\delta_\alpha^\delta \delta_{\gamma}^{\beta},\label{tauid1}
\\
\bar\tau^{(\mu}\tau^{\nu)}=&\tau^{(\mu}\bar\tau^{\nu)}=\delta^{\m\n}\mathds{1}.&&\label{tauid2}
\end{align}
Similarly it will be useful to collect the background fields as
\begin{equation}
(A_A)=(A_{\m a})=(A_{ia},-f_a-U_a)\,,
\end{equation}
This extended background gaugefield has a natural field strength
\begin{equation}
F_{\m a\,\n b}=\partial_{\m a}A_{\n b}-\partial_{\n b}A_{\m a}.\label{asdp}
\end{equation}
Already at this level we see some of the elegance of this new formulation, the supersymmetry conditions \eqref{UAids} become an anti-selfduality condition on this fieldstrength:
\begin{equation}
\frac{1}{2}\epsilon_{\m\n\r\s}F_{\r a\, \s b}=-F_{\m a\, \n b}.\label{asdc}
\end{equation}
The next step is to redefine the fermions, instead of working with a complex doublet, it will be useful to work with 4 real fermions. 

We start by introducing a constant complex 2-vector $\kappa_\alpha${}\footnote{A similar object is discussed in section 4.2.1 of \cite{Moore:2015szp}, where it has its origins in the broken susies from a 4d BPS perspective.}. Note that $\kappa$ is {\it not} Grassmann valued. It is convenient to normalize $\kappa$:
\begin{equation}
\bar{\kappa}\kappa=\bar{\kappa}^\alpha \kappa_\alpha=1.
\end{equation}Now define
\begin{equation}
\chi^{\m a}=\frac{1}{\sqrt{2}}\left(\bar \kappa \bar\tau^\mu \lambda^a+\bar \lambda^a \tau^\mu \kappa\right).\label{chidef}
\end{equation}
Note that $\chi$ is real and Grassmann valued. The key point is that the above is just a field redefinition since it has the inverse:
\begin{equation}
\lambda^a=\frac{1}{\sqrt{2}}\tau^\mu\kappa\,\chi^{\m a},\qquad \bar\lambda^a=\frac{1}{\sqrt{2}}\bar\kappa\bar\tau^\mu\,\chi^{\m a}.\label{lamredef}
\end{equation}

In these new variables also the supersymmetry transformations will take a slightly different form -- see (\ref{susyHKT1}, \ref{susyHKT2}) below -- and it will be useful to redefine the susy parameters as well:
\begin{equation}
\zeta^\mu=\frac{i}{\sqrt{2}}(\bar{\kappa}\bar \tau^\mu\xi-\bar\xi\tau^\mu\kappa )\,,\qquad \xi=-\frac{i}{\sqrt{2}}\tau^\mu\kappa \,\zeta^\mu\,,\qquad \bar{\xi}=\frac{i}{\sqrt{2}}\bar \kappa\bar{\tau}^\mu\,\zeta^\mu.\label{susyredef}
\end{equation}

The detailed redefinitions above can be summarized as
\begin{eqnarray}
 (3,2,1) \mbox{ multiplet} & & (4,4,0) \mbox{ multiplet}\nonumber\\
 (x^{ia},\lambda^{a},D^a) \quad&\qquad\leftrightarrow\qquad & \qquad(x^{\m a},\chi^{\m a})\\
 &\mbox{susy parameters} &\nonumber\\
 \xi &\qquad\leftrightarrow\qquad & \zeta^\m.
\end{eqnarray}

\subsection{Lagrangian and geometry}
Via the field redefinition of the previous subsection the Lagrangian \eqref{Lag1} takes the form
\begin{equation}
L=\underset{L^{(0)}+L^{(1)}}{\underbrace{A_A \dot x^A-\frac{i}{2}F_{AB}\chi^A\chi^B}}+\underset{L^{(2)}}{\underbrace{\frac{1}{2}G_{AB}\left(\dot x^A \dot x^B+i\chi^A \check D_t \chi^B\right)-\frac{1}{12}\partial_{[A}C_{BCD]} \chi^A\chi^B\chi^C\chi^D}}\label{HKTlag}
\end{equation}
where $A=\m a$, $\m=1,2,3,4$, $a=1,\ldots,n$, and
\begin{eqnarray}
G_{\m a\, \n b}&=&\delta_{\m\n}G_{ab}\label{metdef}\\
C_{\m a\, \n b\, \r c}&=&\partial_{\l a}G_{bc}\,\epsilon_{\l\m\n\r}\label{Cdef}
\end{eqnarray}
Note that by construction $\partial_{4a}G_{bc}=0$, which together with the identity \eqref{symmK} and definition \eqref{Hessepot} implies the important symmetry property
\begin{equation}
\partial_{\m a}G_{bc}=\partial_{\m (a}G_{bc)}.\label{Gprop}
\end{equation}
This in turn guarantees that $C_{ABC}$ is totally anti-symmetric and hence defines a 3-form, which provides the torsion of the covariant derivative $\check{D}$ defined by the connection\footnote{This connection is the Bismuth connection, the unique connection compatible with a hermitian structure and totally anti-symmetric torsion. As can be seen in \eqref{whktc} in the current model the three hermitian structures share the same Bismuth connection, one of the requirements for HKT geometry.}
\begin{equation}
	\check{\Gamma}_{AB}^C=\Gamma_{AB}^C+\frac{1}{2}G^{CD}C_{DAB}
\end{equation}
where $\Gamma_{AB}^C$ is the Levi-Civita connection of the metric \eqref{metdef}. Explicitly
\begin{equation}
\check D_t \chi^A=\dot \chi^A+\check \Gamma_{BC}^A\dot x^B\chi^C.
\end{equation}

The Lagrangian \eqref{HKTlag} is invariant under the 4 supersymmetry transformations
\begin{eqnarray}
\delta x^A&=& -i\zeta^\rho ( J^\rho)^A{}_B \chi^B, \label{susyHKT1}\\
\delta \chi^A&=& \zeta^\rho (\bar J^\rho)^A{}_B \dot x^B .\label{susyHKT2}
\end{eqnarray}
These can be directly obtained from the susy transformations \eqref{comptransf} through the field redefinitions of the previous subsection. Here $(J^\rho)=(J^i,\mathds{1})$, $(\bar J^\rho)=(-J^i,\mathds{1})$ in terms of the quaternionic structure
\begin{equation}
(J^i)^{\m a}{}_{\n b}=\delta^a_b (j^i_+)_{\m\n}\label{qs}
\end{equation}
with $j_+^i$ the self-dual quaternionic structure on $\mathbb{R}^4$, see appendix \ref{R4ap} for a precise definition and our conventions. This quaternionic structure appears after the field redefinition through the identity
\begin{equation}
(J^\rho)^{\mu a}{}_{\nu b}=\frac{1}{2} (\bar{\kappa}\bar{\tau}^\nu{\tau}^\mu\tau^\rho\kappa+\bar{\kappa}\bar{\tau}^\rho\bar{\tau}^\mu\tau^\nu\kappa).\label{qstruct}
\end{equation}

As we already alluded to above, supersymmetric invariance of the zeroth and first order part of \eqref{HKTlag} is guaranteed by the anti-self duality property \eqref{asdp}, as is the case in generic ${\mathbf{(4,4,0)}}$ models \cite{Ivanov:2003nk}.

Supersymmetric invariance of the second order part of the Lagrangian \eqref{HKTlag} can be directly related to the tensors $(G,C,J^i)$ defining a (weak\footnote{Here weak refers to the fact that the torsion 3-form $C$ is not closed. Strong HKT geometry requires closure of $C$. In the special case $C=0$ HKT geometry becomes hyperk\"ahler geometry.}) hyperk\"ahler with torsion (HKT) geometry \cite{Coles:1990hr}. We refer to \cite{Fedoruk:2014jba} for a pedagogic and detailed review of HKT geometry and its relation to ${\cal N}=4$ sigma models, for our purposes here we can restrict attention to the following sufficient conditions on the tensors defining a HKT geometry:
\begin{eqnarray}
J^i J^j&=&-\delta^{ij}+\epsilon^{ijk}J^k\qquad \mbox{(quaternion algebra)}\nonumber\\
G_{C(A}(J^i)^C{}_{B)}&=&0\qquad\quad\qquad\qquad \mbox{(hermiticity)}\label{whktc}\\
\check{\nabla}_A (J^r)^B{}_{C}&=&0\qquad\quad\qquad\qquad\mbox{(torsional covariant constant)}.\nonumber
\end{eqnarray}
One can verify by direct computation that the tensors (\ref{metdef}, \ref{Cdef}) and \eqref{qs} satisfy the conditions above. Note that all three complex structures should also be integrable, something which is trivial in our case, since they take constant values in the coordinates we are using, see \eqref{qs}.

So far we have focused on the supersymmetric invariance of the theory, but it is furthermore also conformally invariant.  Since there is a well-studied class of HKT sigma models with ${\cal N}=4$ superconformal symmetry  \cite{Michelson:1999zf} one might expect the conformal quiver quantum mechanics theories we described here to fall into that class. Somewhat surprisingly this is not the case and although, even in HKT form, our models are fully $D(2,1;0)$ invariant they are so in a slightly different and less manifest way than the models discussed in \cite{Michelson:1999zf}. This is a direct consequence of $D^a$ transforming as a primary field, i.e. $\delta D^a=2\dot P D^a+P\dot D^a$ as in \eqref{comptrans}, so that to preserve conformal invariance under the replacement $D^a\rightarrow \dot x^{4a}$ as the first step in figure \ref{reffig} we get the transformation $\delta \dot x^{4a}=2\dot P \dot x^{4a}+P\ddot x^{4a}$ which in turn implies the somewhat peculiar conformal transformations\footnote{Note that strictly speaking the ungauged sigma model is only a formal substep and that our ${\mathbf{(3,4,1)}}$ model is really equivalent to a gauged ${\mathbf{(4,4,0)}}$ sigma model. In particular the precise identification is $D^a\leftrightarrow D_t x^{4a}$ and so conformal invariance only demands $D_t x^{4a}$ to transform as a primary, not necessarily $\dot x^{4a}$. This observation allows for a transformation where $x^{4a}$ remains a primary but we give the gauge field a non-trivial transformation:
	\begin{equation}
	\delta x^{4a}=\dot P x^{4a}+P\dot x^{4a},\qquad \delta B^a=2\dot P B^a+P\dot B^a-\ddot P x^{4a}.\label{nonloctransfo}
	\end{equation}
	This approach however has the disadvantage that we start from a ${\mathbf{(4,4,0)}}$ sigma model that is not conformally invariant but only becomes conformally invariant upon gauging. Using the transformation rule \eqref{nonloctransfo} makes the HKT model conformally invariant even before gauging, be it in a non-standard way.}
\begin{equation}
\delta x^{4a}=\dot P x^{4a}+P\dot x^{4a}-\int_{t_0}^t \ddot P x^{4a} dt'.
\end{equation}
The first two terms coincide with the transformation of a primary field of weight 1, like the other coordinates $x^{ia}$ and which is the transformation assumed in \cite{Michelson:1999zf}. There appears here however the extra third term, which is non-local. Note that this extra term vanishes for the time translations and conformal rescalings, but is present for the special conformal transformation. At a technical level this implies that the HKT metric $G_{AB}$, see \eqref{metdef}, still has a conformal Killing vector, $K^{A}=-2x^{A}$, but that this conformal Killing vector is no longer exact, which is a requirement in the models of \cite{Michelson:1999zf}.

\subsection{Supercharges}
The redefinition of the supersymmetry parameters suggests the redefinition of the charges \eqref{QS} as $Q^\mu=-\frac{1}{\sqrt{2}}(\bar{\kappa}\bar{\tau}^\mu\bar Q+Q\tau^\m \kappa)$. Using the other field redefinitions of section \ref{fredef} the redefined charges can be put into the form
\begin{eqnarray}
Q^4&=&\chi^A\left(\tilde p_{A}-A_{A}-i\omega_{A\,CD}\chi^C\chi^D+\frac{i}{6}C_{ACD}\chi^{C}\chi^{D}\right),\\
Q^i&=&\chi^B (J^i)^{A}{}_B\left(\tilde p_{A}-A_{A}-i\omega_{A\,CD}\chi^C\chi^D+\frac{i}{2}C_{ACD}\chi^{C}\chi^{D}\right).
\end{eqnarray}
These match with the supercharges derived on general grounds in \cite{Smilga:2012wy}. Note that $\omega_{A\, BC}=\omega_A{}^{\underline{DF}}E^{\underline{D}}_BE^{\underline{F}}_C$ with  $E_A^{\underline{B}}$ and $\omega_A{}^{\underline{BC}}$ the vielbien and spin connection associated to the metric \eqref{metdef}, in particular $E^{\underline{\n b}}_{\m a}=\delta^{\n}_\m e^{\underline{b}}_a$, with $e^{\underline b}_a$ the vielbein of the metric\footnote{We should point out that due to the gauge symmetry that removes the overall translational degree of freedom $G_{ab}$ is only semidefinite, with its (projective) inverse defined in \eqref{invdef}. Vielbeine can however still be defined as follows: $G_{ab}=e_a^{\underline c}e_b^{\underline{d}}\delta_{\underline{cd}}$ where $\underline {c},\underline{d}=1,\ldots N-1$. Furthermore we can define their inverses as $G^{ab}=e^a_{\underline c}e^b_{\underline{d}}\delta^{\underline{cd}}$, so that they satisfy the completeness relations $v_a v^b+e_a^{\underline{c}}e_{\underline{c}}^b=\delta_a^b$ and $e_c^{\underline{a}}e_{\underline{b}}^c=\delta^{\underline{a}}_{\underline{b}}$.} $G_{ab}$.  Finally we should stress that $\tilde p_A$ is the canonical momentum associated to $x^A$ via the Lagrangian \eqref{HKTlag}, {\it while keeping $\chi^{\underline{A}}$ fixed}, rather than $\chi^A$. This momentum is related to the canonical momentum as defined in \eqref{pbos} as
\begin{equation}
\tilde p_{\m a}=p_{\m a}-\frac{i}{2}\partial_{\m a}G_{bc}\bar{\lambda}^b\lambda^c+\frac{i}{2}G_{bd}e^{\underline{c}}_c(\partial_{\m a}e^d_{\underline{c}})\chi^{\n b}\chi^{\n c}.
\end{equation}
Reproducing the expressions of \cite{Smilga:2012wy} is interesting, since it paves the way to a quantization via differential forms and an interpretation of supersymmetric groundstates in terms of cohomology.

\section{Discussion and outlook}\label{discsec}
In this work we explicitly exhibited the superconformal symmetry of the Coulomb branch quiver mechanics of D-brane systems with an arbitrary number of centers in an $AdS_2$ scaling limit.
Besides providing explicit examples of multi-particle $D(2,1;0)$-symmetric quantum mechanics specified by scaling quiver data, which is of some interest in itself, it is our  hope that 
our analysis provides a starting point for addressing some conceptual issues in black hole physics. In general, the supersymmetric quantum ground states of the quiver theory describe BPS bound states of D-branes, and when the total charges correspond to those of a black hole it would be of great interest to determine if the quiver theory captures some of the black hole microscopics. In particular, without the scaling limit it is understood how the Coulomb branch of the quiver quantum mechanics corresponds to multi-centered supergravity configurations, while extra states on the Higgs branch are assumed to describe single center black hole microstates \cite{Denef:2007vg, Bena:2012hf, Lee:2012sc, Manschot:2012rx, Lee:2012naa, Messamah:2020ldn}. This suggests that the scaling/AdS$_2$ limit considered in this paper that zooms in on that part of the Coulomb branch that connects to the Higgs branch might be an interesting regime of relevance.

The quantization of the system is beyond the scope of the current work\footnote{The classic reference on quantization of conformally invariant quantum mechanics is \cite{deAlfaro:1976vlx}. Quantization of a class of models with $D(2,1;\a )$ symmetry appears in \cite{Cunha:2016fnr}.}, but let us  say a few words about the space of classical ground states of our Coulomb quiver  mechanics models.
Classical ground states  are configurations  with zero velocity and vanishing fermions,
$\dot x^{ia}= \l^a= \bar \l^a=0$, which solve the D-term constraints
\be
U_a =\sum_{b, b \neq a} {\k_{ab}\over 2 r_{ab}} = 0.\label{classGS}
\ee
The classical moduli space is therefore parametrized by those position vectors  $x^{ia}$, modulo an overall translation, that solve $U_a=0$ and the corresponding solutions preserve all four Poincar\'e
supersymmetries and have vanishing energy and $su(2)$ R-charge. The set of ground states is invariant under rescalings \be x^{ia}\rightarrow \lambda x^{ia},\label{rescaling}\ee for any positive parameter $\l$, and is hence necessarily non-compact\footnote{Note that for non-scaling solutions there exists a natural symplectic form on the space of groundstates \cite{deBoer:2008zn} and that those spaces of groundstates have finite symplectic volume. In the AdS$_2$/scaling limit we consider here the symplectic form vanishes on the space of groundstates, identifying it as a sub-configuration space.}. Since individual ground state solutions are not invariant under rescalings the dilatation symmetry is  spontaneously broken. In contrast the special conformal symmetry, generated by $K$, does not act properly on the moduli space since it generates non-zero velocity, in particular this implies that the set of classical ground states is {\it not} a collection of $SL(2,\mathbb{R})$ orbits. Note that this happens because $K$, although being a symmetry, does not commute with the Hamiltonian\footnote{Of course also $D$ does not commute with the Hamiltonian when acting on generic solutions, but when restricted to the space of groundstates the commutator, which is the Hamiltonian, vanishes.}. It might be interesting to point out that one can obtain a subsector of the theory that is closed under $SL(2,\mathbb{R})$ transformations by allowing only those non-zero velocities that are tangent to the moduli space. Working out the corresponding sigma-model could be a possible future direction.

We now relate these observations to properties of the corresponding  supergravity solutions that were observed in \cite{Bena:2018bbd,Mirfendereski:2018tob}. As was discovered in \cite{Denef:2002ru}, there is a one-to-one correspondence between the classical moduli space and the space of multi-center supergravity solutions. This is because the D-term constraints (\ref{classGS}) precisely coincide with the Denef equations \cite{Denef:2000nb} which govern the existence of the supergravity solution. In the case at hand every solution $x^{ia}$ of (\ref{classGS}) determines a supergravity solution with vanishing angular momentum which is constructed from  a set of harmonic functions 
\be 
H  = \sum_a {\G_a\over |\vec x - \vec x_a|}.
\ee 
The geometry at large radius $r$ takes the form of $AdS_2 \times S^2$
plus corrections. The latter can be systematically derived
from the multipole expansion of the harmonic functions
\be 
H = { \G_{tot}\over r} +\sum_{l=1}^\infty \calo_{i_1 \ldots i_l}^{(l)} {x_{i_1 \ldots i_l} \over r^{2l+1}} \label{multipole}
\ee
where
\be
 \calo_{i_1 \ldots i_l}^{(l)}\sim \sum_a \G_a \left( x^a_{i_1} \ldots  x^a_{i_l} - ({\rm traces})  \right)
\ee
are symmetric, traceless polynomials in the center positions.  The first term in (\ref{multipole}) gives rise to the exact $AdS_2 \times S^2$ throat of a single-centered black hole with charge $\G_{tot}$. The coefficients of correction terms to this throat geometry can be identified with VEVs of the operators 
 $ \calo_{i_1 \ldots i_l}^{(l)}$ in the Coulomb branch quantum mechanics.
A careful analysis \cite{Mirfendereski:2018tob} shows that the leading large $r$ correction to the metric is actually of order 1 and represents a  rotation of
 the $S^2$, proportional to the magnitude of the spin-1 operator
\be 
K^i = \half \sum_{a<b} \k_{ab} x^{iab}.
\ee
Similar examples of `hair' on asymptotically AdS$_2$ backgrounds were discovered in \cite{Bena:2018bbd}.
In the limit $\l \to 0$ in (\ref{rescaling}), where conformal invariance is restored, the supergravity background becomes exactly that of the single-centered black hole. At this point, the Coulomb branch considered in this work matches on to the Higgs branch of the quiver mechanics. 

The above classical picture of spontaneous symmetry breaking on the Coulomb branch and corrections to the supergravity background resembles closely the well-studied holographic description of the Coulomb branch of $\caln=4$ super Yang-Mills \cite{Kraus:1998hv,Klebanov:1999tb}. In the present context there is a caveat however, since here the field theory is a quantum mechanical system and the Coleman-Mermin-Wagner theorem would suggest that the classical  symmetry breaking cannot persist in the quantum theory. 
A related open question is whether the AdS$_2$ scaling  limit of  the quiver quantum mechanics captures some of the black hole microstates and, if so, how the conformal quiver mechanics is related to the putative CFT$_1$ dual to the AdS$_2$ black hole throat. While we  leave  these  interesting issues for further study we  offer here just some remarks.

Since the number of degrees of freedom also goes to infinity in the large charge limit in which supergravity is reliable, it not clear to us if a scenario where  the spontaneous conformal    symmetry breaking persists in the quantum theory can be completely ruled out.
Such a spontaneous  symmetry  breaking 
in the ground states would however be  hard to reconcile with   the standard picture \cite{Sen:2008yk,Maldacena:2016upp} of the CFT$_1$ as a topological theory of singlet ground states.  Similar observations were made in \cite{Bena:2018bbd}.

Another possibility is that conformal symmetry is unbroken and  quiver theory contains normalizeable  ground states which are $D(2,1;0)$ singlets. Since classically the conformally invariant point occurs where the Coulomb and Higgs branches meet, 
an accurate description of these states may have to incorporate the Higgs branch degrees of freedom. In any case, such a scenario would resemble more closely the standard picture of the dual CFT$_1$. A possible relation between quiver quantum mechanics and CFT$_1$  could be that the various quiver theories arising from  different decompositions of the total charge describe superselection sectors of the CFT$_1$.

A related interesting avenue would be to explore controlled deviations from the BPS limit in the quiver mechanics and potential links with nAdS$_2$/nCFT$_1$ \cite{Maldacena:2016upp}.



\section*{Acknowledgements}
We would like to thank Dionysios Anninos, Tarek Anous and Tom\'{a}\v{s} Proch\'{a}zka for interesting discussions. DM and DVdB are supported by TUBITAK grant 117F376. The research of JR was supported
by ESIF and
MEYS (Project CoGraDS - CZ.02.1.01/0.0/0.0/15 003/0000437).

\appendix

	\section{Spinor conventions}\label{Appspinors}
Let $\psi_\a, \a = 1,2$ be an anticommuting, complex, 2-component spinor. Its complex conjugate is denoted as $(\psi_\a)^* \equiv \bar \psi^\a$.
	We  follow the index convention of \cite{Denef:2002ru} where
	we don't define index-lowering or raising operations, rather unbarred spinors $\psi_\a$ always have indices down, barred spinors $\bar \psi^\a$  have indices up and we write insertions of the $SU(2)$ invariant $\e$-tensor explicitly. Pauli matrices $\s^{i \ \b}_{\a}$ always have the first index down and the second one up. Some properties and definitions are
	\bea
	 \bar \l \psi &\equiv& \bar \l^\a \psi_\a,\\
	 	\psi \e \l& \equiv&\psi_\a  \e^{\a\b}\l_\b, \qquad \bar \psi \e \bar \l  \equiv \bar \l^\a \e_{\a\b}\bar \psi^\b ,\\
	  (\psi_\a \l_\b)^* &\equiv& \bar \l^\b \bar \psi^\a,\\
	 		\e_{\a\b} &=& - \e_{\b\a}, \qquad\e_{\a\b}\e^{\b \g} = \d_\a^\g, \qquad \e^{12}=\e_{21}=1,\\
	 		(\s^i \s^j)^{ \ \b}_{\a} &=& \d^{ij}\d_\a^\b+ i \e^{ijk} \s^{k \ \b}_{\a}.
	\eea
	\section{More on the superspace formulation}\label{Appsusy}
		In this Appendix we review the superspace formulation of $\caln =4$ supersymetric quantum mechanics with vector multiplets, referring to   \cite{Ivanov:2002pc,Diaconescu:1997ut} for  more details.

The $d=1,\ \caln =4$ superspace is parametrized by the time $t$ and an anticommuting $su(2)$ spinor doublet $\theta_\a$ together with its complex conjugate  complex conjugate $\bar \theta^\a = (\theta_\a)^*$. 	Supersymmetry generators and supercovariant derivatives are defined as
	\bea
	Q^\a &=& {\pa \over \pa \theta_\a} + i \bar \theta^\a \pa_t, \qquad \bar Q_\a =  {\pa \over \pa \bar \theta^\a} + i  \theta_\a \pa_t ,\\
		D^\a &=& {\pa \over \pa \theta_\a}- i \bar \theta^\a \pa_t, \qquad \bar D_\a =  {\pa \over \pa \bar \theta^\a}  - i  \theta_\a \pa_t,
	\eea
	and satisfy the anticommutation relations
	\be
	\{ Q^\a, \bar Q_\b \} = 2 i \d^\a_\b \pa_t, \qquad 	\{ D^\a, \bar D_\b \} =- 2 i \d^\a_\b \pa_t,
	\ee
	with all other anticommutators vanishing.

	The off-shell vector multiplet can be described by a real scalar superfield 
\be 
V = V^*,
\ee
subject to the gauge equivalence\footnote{$V$ has an additional non-chiral gauge invariance $V\to V + (\overline D^\a D_\a- D_\a \overline D^\a ) \s$.}
\be 
V \sim V + \L + \bar \L,\label{VgaugeApp}
\ee 
where $\L$ is a chiral superfield with $ \overline D_\a \L= D^\a \bar \L=0$.

 From $V$ we can form a triplet of gauge-invariant superfields
\be
\F_i = \half D \s_i   \bar D V 
, \qquad i = 1,2,3.\label{FitoVApp}
\ee
The $\F_i$ satisfy a number of constraints as a consequence of   this relation:
\bea
0&=&D^\a D^\b \F^{\ \g}_\b = \bar D_\a \bar D_\b \F_\g^{\ \b},\label{appconstr1}\\
0&=& \bar D_{(\a} (\F \e)_{\b \g)} = D^{(\a} (\e \F)^{\b\g)},\label{appconstr2}
\eea
where we defined
\be
\F_\a^{\ \b}\equiv \F_i (\s_i)_\a^{\ \b}= \left(\bar D_\a D^\b - \half \d_\a^\b \bar D_\g D^\g\right)V .
\ee
The following useful identities are corollaries of (\ref{appconstr2}): 
\bea
0 &=& D\e D |\F|^{-1} = \overline D \e \overline D |\F|^{-1}, \label{WZWprop1}\\
0 &=& D^\a \left( { \F_\a^{\ \b} \over |\F|^3 }\right)= \overline D_\b \left( { \F_\a^{\ \b} \over |\F|^3 }\right), \label{WZWprop2}
\eea
where $|\F| \equiv (\F_i \F_i )^\half$.

 Parametrizing $V$ as
\bea 
V &=& v + \bar \psi^\a \theta_\a + \bar \theta^\a \psi_\a +z \theta \e \theta  +\bar z   \bar \theta \e \bar \theta - x_i \bar \theta \s_i \theta + C_t \bar \theta \theta\label{appVcomp}\\ && - i  \theta \e \theta  \bar \theta \e \left( \bar \l + \half \bar {\dot \psi} \right)+ i   \left(  \l + \half {\dot \psi} \right) \e \theta \bar  \theta \e \bar \theta + \half \left( D + {i \over 2} \dot C_t -\half \ddot v\right)  \theta \e \theta \bar  \theta \e \bar \theta \nonu \nonumber
\eea
 we obtain the  component
expansion of  $\F_i$
\be
\F_i = x_i + i \bar \l^a \s_i \theta - i \bar \theta \s_i \l+ \bar \theta \s_i \theta D- \e_{ijk}\bar \theta \s_j \theta \dot x_k +\half \bar \theta \e \bar \theta \theta \e \s_i \dot\l -\half  \theta \e \theta \bar \theta  \s_i \e \dot{\bar \l} +{1\over 4} \theta \e \theta  \bar \theta \e \bar \theta \ddot x_i.
\ee

The theories we will consider contain $N$ such vector multiplets (one for each D-brane center) $\F^a_i$ labelled by an  additional index $a= 1, \ldots , N$. A supersymmetric action which is second order in time derivatives  can be obtained from integrating an arbitrary `Hesse potential'  $\calh( \F)$ over superspace. We normalize  the fermionic measure such that
\be
L^{(2)} = \int d^2\theta d^2\bar \theta \calh (\F) \equiv - \left. \calh(\F)\right|_{\theta \e \theta  \bar \theta \e \bar \theta} =  {1 \over 4} \left. \calh(\F)\right|_{\theta_1 \theta_2 \bar \theta^1 \bar \theta^2} .
\ee
Performing the superspace integrals yields
\bea
L^{(2)} &=& {1\over 4} {\pa^2 \calh (x) \over \pa x^a_i \pa x^b_i }\left( \dot x^a_j \dot x^b_j + D^a D^b - i (\dot {\bar \l}^a \l^b - \bar \l^a \dot \l^b)\right)\nonu
&-&  {1 \over 4}  {\pa^3\calh (x) \over \pa x^a_i \pa x^b_i\pa x^c_j } \left(   \bar \l^c\s_j \l^b D^a + \e_{jkl}\bar \l^a \s_k \l^b \dot x^c_l \right) 
- {1 \over 16}  {\pa^4 \calh (x) \over \pa x^a_i \pa x^b_i\pa x^c_j \pa x^d_j }
\l^a \e \l^b \bar \l^c \e \bar \l^d\label{Lsusy1} \\
&+& {1\over 4} {\pa^2 \calh (x) \over \pa x^a_i \pa x^b_j }\left( \dot x^a_i  \dot x^b_j -  \dot x^a_j  \dot x^b_i  - \e_{ijk} (\dot x^a_k D^b - \dot x^b_k D^a + i \dot{\bar \l}^a\s_k \l^b + i \bar \l^a\s_k \dot \l^b  )\right)\nonu
&+&  {1 \over 4}  {\pa^3 \calh (x) \over \pa x^a_i \pa x^b_i\pa x^c_j } \left( \bar \l^a\s_j \l^b D^c -  \bar \l^a\s_j \l^c D^b +  i( \bar \l^a \l^c \dot x^b_k -  \bar \l^c \l^b \dot x^a_k) \right) \nonu
& +& {1 \over 4}  {\pa^3 \calh (x) \over \pa x^a_i \pa x^b_j\pa x^c_k } \left( i \e_{ijk} \bar \l^a \l^b D^c -\e_{ijm}( \bar \l^a \s_k \l^c \dot x^b_m-  \bar \l^c \s_k \l^b \dot x^a_m)\right) \nonu
&-& {1 \over 16}  {\pa^4 \calh (x) \over \pa x^a_i \pa x^b_j\pa x^c_k \pa x^d_l }
\left( 
i\d_{ij}\e_{klm} ( \bar \l^a \e \bar \l^b \l^c \e\s_m\l^d + \bar \l^c \s_m \e \bar \l^d \l^a \e \l^b)
- \e_{ijm}\e_{kln}  \bar \l^a \s_m \e  \bar \l^b  \l^c \e\s_n  \bar \l^d   
\right).\nonu\label{Lsusy2}
\eea
For the models of interest, where the  Hesse potential satisfies
\be 
   {\pa^2 \calh (x) \over \pa x^a_i \pa x^b_j }=  {\pa^2 \calh (x) \over \pa x^b_i \pa x^a_j },\label{symmKapp}
   \ee
 the   terms in (\ref{Lsusy2})  cancel out and only the part (\ref{Lsusy1}) remains.
 
   The Hesse potential $\calh$ is not unique; when (\ref{symmKapp}) holds it is determined up to a `Hessian transformation' 
   \be
   \calh (x) \to \calh (x) + \cals (x),
   \ee
   where $\cals (x)$ satisfies
   \be
   {\pa^2 \cals (x)\over \pa x^a_i \pa x^b_i}=0.\label{Hessetrapp}
   \ee

\section{Useful identities of the coupling functions}
We collect here some useful identitites involving the coupling functions (\ref{U1Xs},\ref{Adefns},\ref{scalingcouplings}) entering in the Lagrangian in the scaling limit:
\bea 
A_{ia} x^{ia} &=&0, \label{id1}\\
\pa_{ia} G_{bc} x^b_j x^c_j &=& - 3  G_{ab} x^b_i, \label{id2}\\
{\pa G_{bc} \over \pa x^{a[i}} x^a_{j]}  &=&0,\label{id3}\\
{\pa U_b \over \pa x^{a[i}} x^a_{j]}  &=&0,\label{id4}\\ 
x^a_i {\pa^2 G_{ab} \over \pa x^c_j x^d_j}& =& - 2  {\pa G_{cd} \over \pa x^b_i },\label{id5}\\
 {\pa A_{ai} \over \pa x^b_j} &=&   {\pa A_{bi} \over \pa x^a_j}.\label{id6}
\eea  
The  identity (\ref{id5}) can be proven using (\ref{symmK},\ref{hompropG}).

\section{Rotational invariance of the magnetic coupling}\label{rotap}
In this Appendix we show that the magnetic coupling $ A_{ia} \dot x^{ia}$ in the Lagrangian  transforms under $su(2)$ R-symmetry  by  a total derivative
\be 
\d_{R^i}(A_{ja} \dot x^a_j) =- {d M_i \over dt}.
\ee
for some functions $M_i$. This is equivalent to the following property of the gauge potential:
\bea 
\d_{R_i} A_{ja} = -\e_{ikl} \pa_{kb} A_{ja}   x^{bl} 
&=& - \e_{ijk} A_{ka} -\pa_{ja} M_{i}.\label{AppRA}
\eea
The functions $M_i$ are determined up to a constant, which can be chosen such that they  satisfy
\be 
\d_{R^i} G_j - \d_{R^j} G_i = - \e_{ijk} G_k.\label{intGs}
\ee
The proof is as follows. The first equality in (\ref{AppRA}) implies that $[\d_i ,\d_j] A_{ka}
= -\e_{ijl} A_{la}$, which using the second equality in (\ref{AppRA}) implies the following condition on the $M_i$:
\be 
\d_{R^i} M_j - \d_{R^j} M_i = - \e_{ijk} M_k + c_{ij}.
\ee
with $c_{ij}$ a constant antisymmetric matrix. Making the redefinition
\be 
M_i \to M_i - \half \e_{ijk} c_{jk}
\ee
the new $M_i$'s satisfy (\ref{intGs}).

An explicit expression for the $M_i$ can be found by realizing  that the Dirac monopole potential $A^{\rm D}_i$  given in (\ref{Adefns}) transforms as a vector under rotations, up to a gauge transformation which rotates the Dirac string direction $n_i$.
Working  out this gauge transformation one finds  that the $M_i$ satisfying (\ref{intGs}) are given by
\be 
M_i = \e_{ijk} A_{aj} x^a_k + U_a x^a_i.
\ee

\section{Primary fields}\label{Appprimary}
The 1-dimensional conformal transformations are defined to act as  PSL$(2,\mathbb{R})$ fractional linear transformations on the time coordinate :
\begin{equation}
t'=\frac{\a t+\b}{\g t+\d}.\label{conffin}
\end{equation}
We define a primary field $\Psi_\D (t)$ with scaling dimension $\D$ to  transform as 
\begin{equation}
{\Psi' _\D}(t')=\left({d t \over d t'}\right)^\D \Psi_\D (t)= (\g t+\d)^{2\D} \Psi_\D (t).\label{prim}
\end{equation}
The infinitesimal version of (\ref{prim}) is
\be 
\d_{\rm conf} \Psi_\D \equiv \Psi_\D'(t) - \Psi_\D (t)= \D\, \dot P\, \Psi_\D + P\, \dot \Psi_\D,
\ee
with $P=u+tv+t^2w$ and $u, v, w$ parametrizing infinitesimal  translations, dilatations and special conformal transformations respectively.

\section{Fermionic derivatives and Poisson brackets}\label{fermder}
In this appendix we shortly list our sign conventions for derivatives and Poisson brackets involving fermionic fields, see also \cite{Casalbuoni:1975bj}. 
\subsection{General considerations}
Let the $\phi^a$ be a collection of fields that can be either Grassmann odd or even, we'll denote the degree as $\epsilon_a$, which is $0$ for bosons and $1$ for fermions. With $F, G$ and $H$ we denote Grassmann valued functions of these fields. Note that since the fields don't commute there will be a difference between the 'right' and 'left' derivative:
\begin{equation*}
\frac{\partial_L}{\partial \phi^a}(F G)=\frac{\partial_L F}{\partial \phi^a}G+(-1)^{\epsilon_F\epsilon_a}F\frac{\partial_L G}{\partial \phi^a}\qquad \frac{\partial_R}{\partial \phi^a}(F G)=(-1)^{\epsilon_G\epsilon_a}\frac{\partial_R F}{\partial \phi^a}G+F\frac{\partial_R G}{\partial \phi^a} .
\end{equation*} 
For example
\begin{equation*}
\frac{\partial_L}{\partial\l_1}(\l_2\l_1)=-\l_2\qquad \frac{\partial_R}{\partial\l_1}(\l_2\l_1)=\l_2.
\end{equation*}
Note that 
\begin{equation*}
\frac{\partial_L F}{\partial \phi^a}=(-1)^{(\epsilon_F+1)\epsilon_a}\frac{\partial_R F}{\partial \phi^a}.
\end{equation*}

Given a Lagrangian for the fields $\phi^a$ one can define the canonical momentum using either the left or right derivative: 
\begin{equation*}
\pi_a^L=\frac{\partial_L L}{\partial\dot \phi^a}=(-1)^{\epsilon_a}\frac{\partial_R L}{\partial \phi^a}=(-1)^{\epsilon_a}\pi_a^R.
\end{equation*}
One is free to choose either momentum, but this choice is correlated with the proper choice of Hamiltonian action:
\begin{equation*}
L=\pi_a^R \dot \phi^a-H=\dot \phi^a\pi_a^L -H.
\end{equation*}
One can then define the following super-Poisson bracket:
\begin{equation*}
\{F,G\}=\frac{\partial_R F}{\partial \phi^a}\frac{\partial_L G}{\partial \pi_a^R}-\frac{\partial_R F}{\partial \pi^L_a}\frac{\partial_L G}{\partial \phi^a}
\end{equation*}
which satisfies 
\begin{equation*}
\{F,G\}=(-1)^{1+\epsilon_F\epsilon_G}\{G,F\}
\end{equation*}
and the super-Jacobi identity. Note that the canonical relations that follow are
\begin{equation*}
\{\phi^a,\pi_b^R\}=\delta_b^a=-\{\pi_b^L,\phi^a\}.
\end{equation*}

One also derives the properties
\begin{equation}
\{F,GH\}=\{F,G\}H+(-1)^{\epsilon_F\epsilon_G} G\{F,H\}\qquad \{FG,H\}=F\{G,H\}+(-1)^{\epsilon_H\epsilon_G} \{F,H\}G
\end{equation}
which when combined lead to
\begin{eqnarray}
\{F_1F_2,G_1G_2\}&=&F_1\{F_2,G_1\}G_2+(-1)^{\epsilon_{F_2}\epsilon_{G_1}}\{F_1,G_1\}F_2G_2\\&&+(-1)^{\epsilon_{F_2}\epsilon_{G_1}}F_1G_1\{F_2,G_2\}+(-1)^{\epsilon_{F_1}\epsilon_{F_2}+\epsilon_{G_1}\epsilon_{G_2}}F_2\{F_1,G_2\}G_1 .\nonumber
\end{eqnarray}

\subsection{Keeping right}
The previous subsection makes it clear one is free to choose either one of left/right, or even work in a mixed formulation. In this paper we will choose to interpret all derivatives with respect to fermionic  fields as right derivatives\footnote{In contrast, for derivatives with respect to the superspace coordinates $\theta_\a, \bar \theta^\a$ e.g. in (\ref{suconfKV}), we follow the standard convention that these act from the left.}, and to ease notation we will drop the superscript $R$. The key identities above then take the form:
\begin{equation*}
\frac{\partial}{\partial \phi^a}(F G)=(-1)^{\epsilon_G\epsilon_a}\frac{\partial F}{\partial \phi^a}G+F\frac{\partial G}{\partial \phi^a} .
\end{equation*} 
For example
\begin{equation*}
\frac{\partial}{\partial\l_1}(\l_2\l_1)=\l_2.
\end{equation*}

Our canonical momentum is defined as:
\begin{equation*}
\pi_a=\frac{\partial L}{\partial\dot \phi^a},
\end{equation*}
and in the Hamiltonian variational principle the Lagrangian is
\begin{equation*}
L=\pi_a \dot \phi^a-H.
\end{equation*}
The super-Poisson bracket is:
\begin{eqnarray}
\{F,G\}&=&(-1)^{\epsilon_a(\epsilon_G+1)}\frac{\partial F}{\partial \phi^a}\frac{\partial G}{\partial \pi_a}-(-1)^{\epsilon_F\epsilon_G}(-1)^{\epsilon_a(\epsilon_F+1)}\frac{\partial G}{\partial \pi_a}\frac{\partial F}{\partial \phi^a}.
\end{eqnarray}
Or more shortly, if we define
\begin{equation}
F*G=(-1)^{\epsilon_a(\epsilon_G+1)}\frac{\partial F}{\partial \phi^a}\frac{\partial G}{\partial \pi_a}
\end{equation}
then 
\begin{equation}
\{F,G\}=F*G-(-1)^{\epsilon_F\epsilon_G} G*F
\end{equation}
so that one sees directly that 
\begin{equation*}
\{F,G\}=-(-1)^{\epsilon_F\epsilon_G}\{G,F\}.
\end{equation*}
Note that the canonical relations that follow are
\begin{equation*}
\{\phi^a,\pi_b\}=\delta_b^a.
\end{equation*}

\subsection{Some useful formulas}
Applied to the systems of interest, where the phase space variables are $(x^{ia},p_{ia}, \l^a_\a, \p_a^\a = i \bar \l_a^\a)$, these conventions lead to the following formulas for the Poisson brackets of arbitrary (even or odd) phase space functions: 
\bea 
\{ B_1 , B_2 \} &=& {\pa B_1 \over \pa x^a_i}  {\pa  B_2 \over \pa p_{ai}}- {\pa   B_1  \over \pa p_{ai}}  {\pa  B_2 \over \pa x^a_i} + i \left(
{\pa B_1  \over \pa \l^a_\a}  {\pa  B_2 \over \pa  \bar \l_a^\a}+  {\pa   B_1  \over \pa  \bar \l_a^\a} {\pa  B_2\over \pa \l^a_\a}\right)\nonu
\{ B , F \} &=&  - \{ F , B \} = {\pa B \over \pa x^a_i}  {\pa F \over \pa p_{ai}}- {\pa  B \over \pa p_{ai}}  {\pa F \over \pa x^a_i} - i \left(
{\pa B \over \pa \l^a_\a}  {\pa F\over \pa  \bar \l_a^\a}+  {\pa  B \over \pa  \bar \l_a^\a} {\pa  F\over \pa \l^a_\a}\right)\nonu
\{ F_1 , F_2 \} &=&  {\pa F_1\over \pa x^a_i}  {\pa F_2 \over \pa p_{ai}}- {\pa  F_1 \over \pa p_{ai}}  {\pa F_2 \over \pa x^a_i} - i \left(
{\pa F_1\over \pa \l^a_\a}  {\pa F_2\over \pa  \bar \l_a^\a}+  {\pa F_1 \over \pa  \bar \l_a^\a} {\pa F_2\over \pa \l^a_\a}\right).\label{PBrules}
\eea 
 It follows that the Poisson  bracket has the reality properties (using that conjugation is defined to  revert the order of Grassmann variables)
\be
\{ \overline{B}_1 ,\overline{B}_2 \} = \overline{ \{ B_1 , B_2 \} }, \qquad
\{ \overline{B} ,\overline{F} \} = \overline{ \{ B , F \} }, \qquad
\{ \overline{F}_1 ,\overline{F}_2 \} = - \overline{ \{ F_1 , F_2 \} }.\label{PBreality}
\ee

\section{$\mathbb{R}^4$, its quaternionic structures and (anti-) self-duality}\label{R4ap}
Let us consider $\mathbb{R}^4$ with coordinates $x^\mu$, $\mu=1,2,3,4$. The metric is the standard Euclidean one, $g_{\m\n}=\delta_{\m\n}$ and for this reason we can be completely careless about raising and lowering indices. By convention we'll take 'all' indices to be lower indices.

Let us start the discussion with anti-symmetric two tensors (or 2-forms) $\omega_{\m\n}=\omega_{[\m\n]}$, we can split the space of these forms into two orthogonal parts by the projectors
\begin{equation}
P_{\m\n\,\r\s}^{\pm}=\frac{1}{4}(\delta_{\m\r}\delta_{\n\s}-\delta_{\m\s}\delta_{\n\r}\pm \epsilon_{\m\n\r\s}),
\end{equation}
i.e. they satisfy
\begin{equation}
P^{\pm}_{\m\n\k\l}P^{\pm}_{\k\l\r\s}=P^\pm_{\m\n\r\s}\qquad P^{\pm}_{\m\n\k\l}P^{\mp}_{\k\l\r\s}=0.\label{projeqs}
\end{equation}
So if we define $\omega_{\m\n}=\omega^+_{\m\n}+\omega^-_{\m\n}$ with $P^{\pm}_{\m\n\r\s}\omega^{\pm}_{\r\s}=\omega^{\pm}_{\m\n}$ then these are the self dual and anti-self dual parts:
\begin{equation}
\frac{1}{2}\epsilon_{\m\n\r\s}\omega^\pm_{\r\s}=\pm \omega^\pm_{\m\n}.
\end{equation}

This natural split into the self dual and anti-self dual subspace is closely related to the two possible quaternionic structures on $\mathbb{R}^4$. Namely, we can define\footnote{Note that these complex structures are essentially equal to the 't Hooft (anti)-self dual symbols: $(j^r_+)_{\m\n}=\eta^r_{\m\n}$, $(j^r_-)_{\m\n}=\bar\eta^r_{\m\n}$}
\begin{equation}
(j_{\pm}^i)_{\m\n}=\mp 4 P^{\pm}_{\m\n\,i4}.	\label{jdef}
\end{equation}
These satisfy the quaternionic algebra:
\begin{equation}
(j_\pm^i)_{\m\r}(j_{\pm}^j)_{\r\n}=-\delta_{\m\n}\delta^{ij}+\epsilon^{ijk}(j_\pm^k)_{\m\n}.
\end{equation}
Note that all complex structures \eqref{jdef} are anti-symmetric, which is equivalent to the metric being Hermitian with respect to all of them:
\begin{equation}
(j^i_\pm)_{\r\m}\delta_{\r\n}=-\delta_{\m\r}(j^i_\pm)_{\r\n}.
\end{equation}
It also follows directly from \eqref{projeqs} that these complex structures are (anti-)self dual:
\begin{equation}
P^\pm_{\m\n\rho\sigma}(j^i_{\pm})_{\rho\sigma}=\pm (j^i_{\pm})_{\m\n}.
\end{equation}

\bibliographystyle{ytphys}
\bibliography{draft_current}
\end{document}